\documentclass[conference,compsoc]{IEEEtran}
\usepackage{amsmath,amssymb,amsfonts}
\usepackage{algorithmic}
\usepackage{graphicx}
\usepackage{textcomp}
\usepackage{xcolor}
\usepackage{url}
\usepackage{latexsym}
\usepackage[backend=biber]{biblatex}
\usepackage{graphicx}
\usepackage{amsmath}
\usepackage{booktabs}
\usepackage{array}
\usepackage{float}
\usepackage[colorlinks=true, linkcolor=blue, citecolor=blue, urlcolor=blue]{hyperref}
\usepackage{booktabs}
\usepackage{makecell}
\usepackage{tikz}
\usetikzlibrary{shapes,arrows,positioning,fit,backgrounds}
\usepackage{booktabs}
\usepackage{makecell}
\usepackage{multirow}
\usepackage{array}
\usepackage[most]{tcolorbox}

\usepackage{tcolorbox}
\tcbset{
  promptbox/.style={
    colback=gray!10,
    colframe=black!50,
    boxrule=0.4pt,
    arc=2pt,
    left=6pt,right=6pt,top=6pt,bottom=6pt,
    fontupper=\normalsize           
  }
}

\usepackage{booktabs,adjustbox,listings,xcolor}
\usepackage{tabularx,makecell,booktabs}
\setcellgapes{3pt}

\usepackage{listings}
\usepackage{xcolor}
\usepackage{tabularx}

\usepackage{pgfplots}
\pgfplotsset{compat=1.18}

\newcommand{\parhead}[1]{\noindent \textbf{#1.}}

\definecolor{codegreen}{rgb}{0,0.6,0}
\definecolor{codegray}{rgb}{0.5,0.5,0.5}
\definecolor{codepurple}{rgb}{0.58,0,0.82}
\definecolor{backcolour}{rgb}{0.95,0.95,0.92}

\lstdefinestyle{mystyle}{
    backgroundcolor=\color{backcolour},   
    commentstyle=\color{codegreen},
    keywordstyle=\color{magenta},
    numberstyle=\tiny\color{codegray},
    stringstyle=\color{codepurple},
    basicstyle=\ttfamily\footnotesize,
    breakatwhitespace=false,         
    breaklines=true,                 
    captionpos=b,                    
    keepspaces=true,                 
    numbers=left,                    
    numbersep=5pt,                  
    showspaces=false,                
    showstringspaces=false,
    showtabs=false,                  
    tabsize=2
}

\def\BibTeX{{\rm B\kern-.05em{\sc i\kern-.025em b}\kern-.08em
    T\kern-.1667em\lower.7ex\hbox{E}\kern-.125emX}}
\begin{document}


\title{NetInspector: Measuring and Improving LLM Capabilities for Reliable Intent-Based Networking Policy Generation}

\author{
\IEEEauthorblockN{Yuxuan Zhang}
\IEEEauthorblockA{
\textit{Texas A\&M University}\\
yuz516@tamu.edu}
\and
\IEEEauthorblockN{Hongxin Hu}
\IEEEauthorblockA{
\textit{University at Buffalo}\\
hongxinh@buffalo.edu}
\and
\IEEEauthorblockN{Guofei Gu}
\IEEEauthorblockA{
\textit{Texas A\&M University}\\
guofei@cse.tamu.edu}
}

\maketitle

\begin{abstract}

Modern networks are large in scale and heterogeneous in configuration, making manual policy management increasingly impractical. Intent-Based Networking (IBN) addresses this by automating the translation of high-level operator goals into low-level network configurations. Yet existing IBN systems rely on static heuristics and fixed-feature classifiers that generalize poorly to distribution shifts such as new service definitions or evolving phrasing in operator requests. This brittleness resulted in security threats where intents that conflict with existing security policies can silently pass the resolution check and reach the network, producing misconfigurations with real operational consequences. Large Language Models (LLMs), with strong reasoning and translation capabilities demonstrated across many domains, are a natural candidate for IBN policy generation. However, it is unclear whether LLMs can be reliably applied to this task, nor whether their use mitigates or worsens the underlying security risk.

In this work, we bridge this gap with a systematic measurement study of LLMs in the IBN policy generation pipeline. Our study shows that while fine-tuned LLMs excel at intent translation, they exhibit false negative rates of 35–48\% when checking whether a proposed intent violates an existing security policy. The root cause is not a lack of logical reasoning capability, but a lack of persistent grounding in network topology and group hierarchy: LLMs cannot reliably navigate the large network state to retrieve the facts their reasoning depends on. Motivated by this finding, we introduce NetInspector, a three-layer agentic framework that enforces a verify-then-act protocol, decoupling information retrieval from reasoning so that the LLM focuses on symbolic reasoning while every policy decision is grounded in verifiable network facts retrieved from a live Environment Layer before approval. On NetInspector-Bench, a 2,224-sample synthetic benchmark spanning campus, enterprise, and WAN topologies, NetInspector reduces FNR by over 30\% relative to ungrounded baselines and remains robust under linguistic distribution shifts.

\end{abstract}



\section{Introduction}

The increasing scale and heterogeneity of modern networks have made manual configuration a primary source of security vulnerabilities\cite{facebookoutage,cnnfacebookoutage,gartnerdatareport}. Traditional management relies heavily on human experts to translate high-level network management and security goals (e.g., isolate guest traffic) into low-level enforcement primitives (e.g., ACLs, VLAN tags). This manual workflow is not only time-consuming but fundamentally error-prone; a single misconfiguration can lead to open attack surfaces, routing loops, or service outages~\cite{cnnfacebookoutage}. To mitigate these risks, Intent-Based Networking (IBN) has emerged as a paradigm shift, moving the control plane from imperative configuration to declarative policy management. By allowing operators to specify ``what'' is needed rather than ``how'' to implement it, IBN promises to automate the enforcement of security invariants and reduce the vulnerabilities caused by human error.

\begin{figure}[ht]
	\centering
	\includegraphics[width=0.5\textwidth]{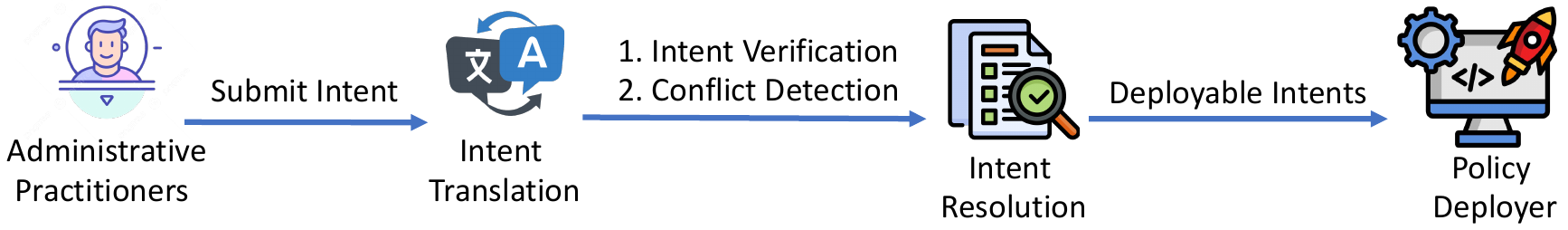}
	\caption{Illustration of the IBN Policy Generation process}\label{fig:IBN-lifecycle}
\end{figure}

However, the security assurance of an IBN system is entirely dependent on the integrity of its policy generation pipeline. As defined by Leivadeas et al.~\cite{leivadeas2022survey}, realizing IBN involves complex translation stages, converting natural language intent into formal specifications (e.g., Nile~\cite{Wool2011RuleComplexity}) before deployment. As illustrated in Figure~\ref{fig:IBN-lifecycle}, this process hinges on two critical functions: Intent Translation (parsing the request) and Intent Resolution (verifying the requested policy against the existing security policies in deployment). While Translation ensures the system understands the user, Intent Resolution acts as the security enforcement layer---the last automated checkpoint before deployment---responsible for blocking any intent that would violate a standing security policy. A missed violation can silently bypass an access restriction, redirect traffic across untrusted segments, or over-commit shared resources, and the resulting misconfigurations can lead to real operational and security incidents in production networks~\cite{facebookoutage}. If this stage fails, the automation engine becomes a direct vector for committing unsafe policy changes to the network.

\begin{figure}[ht]
	\centering
	\includegraphics[width=0.5\textwidth]{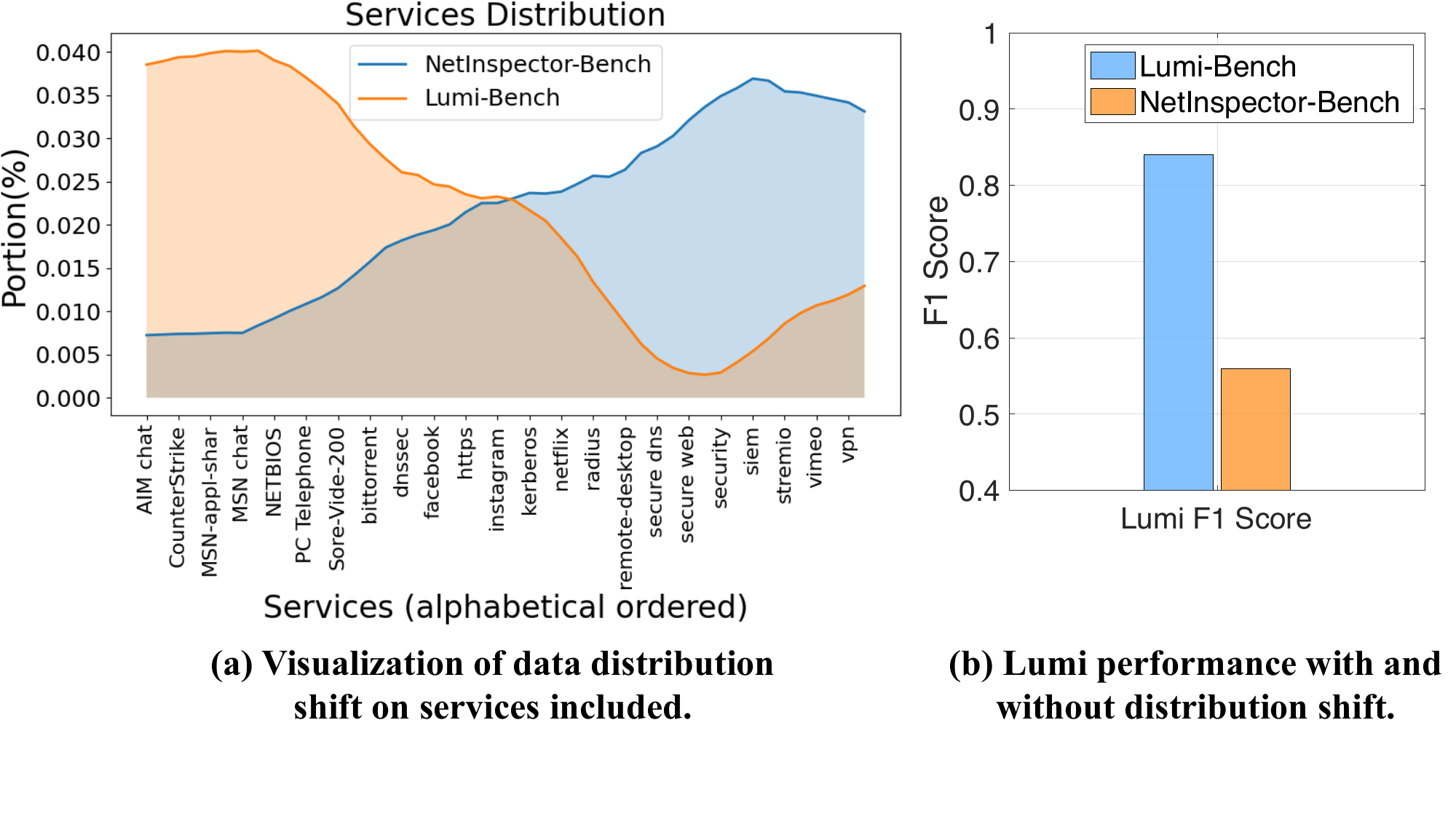}
	\caption{Existing state-of-art approach (Lumi\cite{jacobs2021hey}) shows limited robustness under data distribution shift. Figure(a) shows the distribution shift between our benchmark and Lumi's benchmark. Figure(b) presents Lumi's performance in F1 score on both datasets.}
     \label{fig:motivation}
\end{figure}

To secure this pipeline, prior work has relied largely on static heuristics and rigid rule-based parsers. In Intent Translation, systems typically extract entities into predefined templates using regular expressions~\cite{tuncer2018northbound}. For Intent Resolution, conflict detection often depends on extracting static numerical features to train classifiers. While early Machine Learning (ML) approaches~\cite{jacobs2018refining, yang2020intent} and RNN-based models proposed by Jacobs et al.~\cite{jacobs2021hey} improved flexibility, they heavily rely on models trained on fixed feature sets. From a security perspective, this reliance on static heuristics represents a significant vulnerability: these models are brittle and fail to generalize under distribution shifts. As shown in Figure~\ref{fig:motivation} (a), a shift in service definitions or phrasing, common in evolving network environments, can render these static detectors obsolete. Figure~\ref{fig:motivation} (b) demonstrates the consequence: the failure to detect such conflicts increases drastically, allowing unsafe intents to bypass the resolution check and reach the network as real misconfigurations. To overcome this brittleness, a growing body of work~\cite{dzeparoska2023llm, dzeparoska2023policygen, mekrache2024llmconfig, dzeparoska2024drift, habib2024rl} has explored Large Language Models (LLMs) as policy generators in IBN, leveraging their semantic flexibility to map diverse natural-language intents into structured policies without rigid templates. These efforts, however, primarily showcase that LLMs are capable of producing syntactically valid policies; they do not investigate whether the policies LLMs generate can reliably avoid violating any existing security policies---the central concern once LLMs are placed on the security enforcement path.

In this paper, we seek to address the central research question: \textit{Can LLMs reliably catch intents that violate an existing security policy before deployment?} To this end, we conduct a systematic measurement study of LLMs across the IBN policy generation pipeline. On a synthetic benchmark constructed from real network topologies, we compare representative prompting and fine-tuning strategies against established non-LLM baselines on both Intent Translation and Intent Resolution. The results show that although fine-tuned LLMs translate natural-language intents into structured policies with high accuracy, their false negative rate on detecting security-policy violations remains at 35--48\%. A closer examination indicates that this failure does not stem from limitations in logical reasoning. Rather, LLMs lack persistent grounding in the network state and cannot reliably navigate the topology and group hierarchy to retrieve the facts on which their reasoning depends.

Building on this insight, we propose \textbf{NetInspector}, a three-layer agentic framework that sits between operator intent and policy deployment. NetInspector enforces a \textit{verify-then-act} protocol that decouples information retrieval from reasoning: a dedicated Environment Layer holds a live, queryable representation of the network state, and the LLM only commits to a policy decision after explicitly retrieving and reasoning over the relevant facts. This decoupling lets the LLM concentrate on the symbolic reasoning it is well-suited for, while removing the burden of navigating the large network state. Across campus, enterprise, and WAN topologies in our benchmark, NetInspector reduces FNR by over 30\% relative to ungrounded baselines and remains robust under linguistic distribution shifts, closing the security gap left by prior approaches.

We summarize our contributions as follows:
\begin{enumerate}
    \item We conduct the first systematic evaluation of LLMs as security-policy checkers in IBN policy generation, revealing that their inability to reason over network topology and group hierarchies produces FNR of 35--48\%, leaving the majority of unsafe intents undetected before deployment.
    \item We propose \textbf{NetInspector}, a three-layer agentic framework built on a \textit{verify-then-act} design principle. By grounding LLM reasoning in verifiable network state, NetInspector reduces the FNR for catching policy-violating intents by over 30 percentage points versus ungrounded baselines, while maintaining robustness under distribution shifts.
    \item We open-source \textbf{NetInspector-Bench}, a 2,224-sample benchmark spanning campus, enterprise, and WAN topologies with six categories of security-policy violations---such as privilege escalation, isolation bypass, and denial of service---enabling reproducible evaluation of security-oriented IBN policy enforcement.
\end{enumerate}


\section{Problem Statement and Related Work}
\label{sec:threat-model}

\parhead{System Model}
We consider an Intent-Based Networking (IBN) system through which network operators (e.g., administrators, lab managers, or service owners) submit natural-language intents $i_\text{new} \in \Sigma^*$ to manage the underlying network, modeled as a state tuple $\sigma = (G, H, c)$ that captures the physical topology $G = (V, E)$, the group-containment hierarchy $H$ (a DAG over user and resource groups), and the link-capacity function $c : E \to \mathbb{R}_+$. A typical intent expresses a high-level goal (e.g., \textit{``allow Engineering hosts to reach the project file server during business hours''}). A translation function $f_T$ maps this utterance into a candidate policy $P_\text{new} = f_T(i_\text{new})$, expressed as a tuple $(s, a, o, \tau)$ encoding a subject, an action (e.g., \texttt{ALLOW}, \texttt{DENY}, \texttt{RATE\_LIMIT}), an object, and a condition such as a temporal scope or priority. Before $P_\text{new}$ is committed to the network as low-level enforcement primitives (ACLs, VLAN tags, quality-of-service reservations), it must pass through the Intent Resolution stage---the last automated checkpoint that verifies whether the requested change is compatible with the standing security-policy library $\mathcal{L} = \{P_1, \dots, P_N\}$ maintained by the security team. From the operator's perspective, the interaction surface is the natural-language intent alone: the burden of materializing $f_T$, accessing $\sigma$, and cross-checking $P_\text{new}$ against the full library $\mathcal{L}$ is delegated entirely to the system.

\parhead{Threat Model}
We assume a \emph{benign but fallible} insider: the operator is a legitimate user whose intent is not necessarily malicious, but whose knowledge of $\mathcal{L}$ and $\sigma$ is necessarily incomplete in any realistically sized deployment. The threats we consider therefore arise not from adversarial intent injection, but from \emph{unintentional conflicts} between $P_\text{new}$ and one or more standing policies---conflicts that the operator is in no position to foresee at submission time. Formally, we say $P_\text{new}$ \emph{violates} $\mathcal{L}$ under $\sigma$ whenever $\exists\, P_i \in \mathcal{L} : \mathrm{Conflict}(P_\text{new}, P_i; \sigma)$, where $\mathrm{Conflict}(\cdot)$ is a class-specific predicate capturing the semantics of a particular violation type. Representative instantiations include privilege escalation~\cite{alshaer2004discovery, yuan2006fireman, valenza2022atomizing, li2023deep}, where an over-broad access rule subsumes a previously restricted subgroup; isolation bypass~\cite{kazemian2012header, khurshid2013veriflow}, where a forwarding rule traverses an untrusted segment; and denial of service~\cite{porras2012security, cinmere2024direct}, where a reservation combined with existing allocations exceeds physical link capacity. The predicate set is open by construction: additional violation classes can be admitted without changing the surrounding framework. When the Intent Resolution stage fails to detect such a violation, the conflicting policy is silently committed to the network. Section~\ref{subsec:case-study} presents a detailed example of each representative class.

\parhead{Detection Objective}
A reliable security-policy verification system for IBN can be modeled as a function $v : (P_\text{new}, \mathcal{L}, \sigma) \to \{\textsc{Conflict}, \textsc{Safe}\}$, and a deployable instance of $v$ must satisfy three requirements. First, it should accept natural-language intents in the form operators actually submit, without forcing them to enumerate or know the contents of $\mathcal{L}$---that is, the system must materialize the translation function $f_T$ end-to-end. Second, every decision must be grounded in $\sigma$---topology, group membership, deployed access rules, and current resource allocations---rather than reasoning over surface syntax alone. Third, $v$ should approach two ideal correctness properties: \emph{soundness}, that $v(\cdot) = \textsc{Conflict}$ implies $\exists\, P_i \in \mathcal{L} : \mathrm{Conflict}(P_\text{new}, P_i; \sigma)$, so that benign intents are not unnecessarily rejected; and \emph{completeness}, that $\exists\, P_i \in \mathcal{L} : \mathrm{Conflict}(P_\text{new}, P_i; \sigma)$ implies $v(\cdot) = \textsc{Conflict}$, so that violating intents are reliably blocked before deployment. These requirements jointly motivate the design of NetInspector, presented in Section~\ref{sec:netinspector}.

\parhead{Related Work} Over the past decade, a wide range of research has investigated how to implement IBN in practice. Existing approaches can be broadly categorized into non-LLM-based methods and LLM-based solutions, each with distinct focuses and limitations.

\parhead{Policy Generation in IBN} Earlier works in IBN primarily relied on rule-based systems, traditional natural language processing, and machine learning techniques. Kim et al. proposed IBCS, an intent-based framework tailored for cloud security services, emphasizing policy translation and automation for secure service provisioning \cite{kim2020ibcs}. Similarly, Han et al. developed an intent-based virtualization platform for SDN, focusing on transforming user-defined intents into virtual network configurations \cite{han2016virtualization}. While these systems provided a degree of automation, they typically required domain-specific expertise and lacked generalizability.
Tuncer et al. introduced a northbound interface for software-defined networks, aiming to bridge the gap between user policies and low-level configurations using predefined templates \cite{tuncer2018northbound}. Jacobs et al. explored refining network intents to adapt to evolving conditions, enhancing system flexibility \cite{jacobs2018refining}. However, these approaches often depended on rigid schemas and static rules, which limited their scalability.

Yang et al. extended IBN concepts to optical networks by incorporating AI for automated operation and maintenance \cite{yang2020intent}, while Savi and Siracusa proposed service-aware provisioning for multi-layer transport networks \cite{savi2018provisioning}. Although effective in specialized domains, these systems did not generalize well across heterogeneous networks.
Li et al. provided a comprehensive survey on deep learning for named entity recognition (NER), a foundational technique in parsing user intents \cite{li2022surveyNER}. Scheid et al. proposed the use of controlled natural languages for intent specification, improving interpretability at the cost of flexibility \cite{scheid2020blockchain}. Singh et al. applied IBN to vehicular edge computing \cite{singh2021vehicular}, and Anand et al. developed an intent-driven telemetry framework for in-band monitoring \cite{anand2018point}. In parallel, Du et al. formalized intent policy formats through the IETF’s ANIMA working group \cite{du2016anima}, contributing to standardization efforts.
Despite these contributions, non-LLM approaches generally require handcrafted rules, offer limited support for ambiguous or diverse inputs, and lack adaptability when applied to new scenarios.

\parhead{LLM in IBN} The advent of large language models (LLMs) has led to new opportunities for automating IBN tasks with greater flexibility. Mekrache and Ksentini proposed an LLM-enabled service configuration framework capable of converting natural language intents into network configurations \cite{mekrache2024llmconfig}. Tu et al. similarly demonstrated the use of LLMs to translate user intent into configuration commands, offering improved handling of diverse input formats \cite{tu2025intentconfig}.
Dzeparoska et al. introduced LLM-based policy generation for IBN \cite{dzeparoska2023policygen}, while Habib et al. incorporated attention-based reinforcement learning with LLMs for network optimization and decision-making \cite{habib2024rl}. Fuad et al. proposed a comprehensive framework integrating LLMs into intent-driven SDN control systems \cite{fuad2024framework}. More recently, Dzeparoska et al. investigated LLM-driven intent assurance guided by intent drift detection \cite{dzeparoska2024drift}.

While these studies have demonstrated the potential of LLMs in enhancing intent translation and assurance, they are generally limited to task-specific evaluations, often focusing on translation alone. They lack a comprehensive analysis of LLM capabilities in reliably perform the tasks across the full IBN policy generation lifecycle, and offer limited guidance on how to design effective LLM prompts, handle edge cases, or interpret LLM output reliability.




\section{Understanding LLM's Capability in Policy Generation}
\label{sec:measurement}
\subsection{Overview}
To systematically explore whether large language models (LLMs) can reliably assist network maintainers in tasks for network policy generation, we further decompose this overarching goal into four focused research questions.
\begin{itemize}
    \item \textbf{Q1: How does LLM perform compared to SOTA approaches in the tasks for policy generation?}
    \item \textbf{Q2: How does prompt engineering affect LLM's performance?}
    \item \textbf{Q3: How does fine-tuning affect LLM's performance?}
    \item \textbf{Q4: In what tasks/scenarios does LLM struggle?}
\end{itemize}

\begin{figure}[ht]
	\centering
	\includegraphics[width=0.5\textwidth]{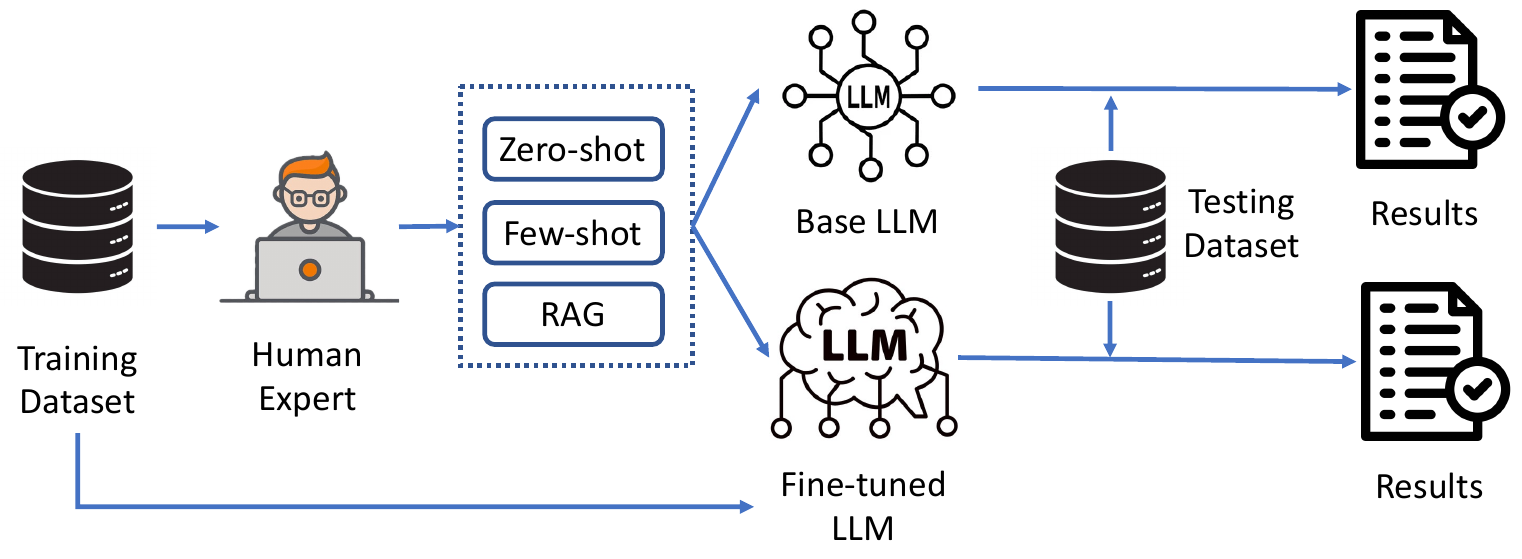}
	\caption{Measurement Study Pipeline}\label{fig:evaluation-pipeline}
\end{figure}

Figure~\ref{fig:evaluation-pipeline} presents the overall workflow of our evaluation. Our evaluation pipeline starts with a carefully crafted benchmark dataset constructed from open-source datasets from previous work\cite{jacobs2021hey}. With our created benchmark, we use a split training set to fine-tune models and build the retrieval base for RAG and few-shot prompting. We design task-specific prompts with a description of the task, step-by-step instructions, and optionally examples. Prompt design details are presented in Table~\ref{tab:prompt_templates}. We assess four model types: zero-shot prompting, few-shot prompting, retrieval-augmented generation (RAG), and fine-tuned LLMs. We compare results from LLMs to the baseline approach using standard evaluation metrics such as accuracy, precision, recall, and F1 score.

\subsection{NetInspector-Bench}
Existing IBN datasets~\cite{jacobs2021hey,wang2024netconfeval} were designed for generic conflict detection and intent translation, not for evaluating security invariant enforcement. They are limited to campus-scale topologies, rely on rigid syntactic templates that allow models to overfit to surface patterns, and do not organize violation types into security threat classes. We constructed NetInspector-Bench to address both gaps — first by extending the coverage and linguistic diversity of prior benchmarks, and second by grounding the evaluation in concrete security threat categories.

\parhead{Extension of Prior Benchmarks}
NetInspector-Bench builds on the violation taxonomy of Jacobs et al.~\cite{jacobs2021hey} and extends it in two dimensions. \textit{Topological coverage}: we expand from campus-only to Campus, Enterprise, and WAN domains using real-world graphs from Topology Zoo~\cite{knight2011internet}, requiring models to reason about path-dependent constraints at different operational scales. \textit{Linguistic diversity}: we inject state-modifying verbs and shuffle clause orders to prevent keyword overfitting, and synthesize scope-shadowing pairs where violation detection hinges entirely on resolving hierarchical group containment.

\parhead{Violation Types}
We include six violation types: \textit{negation} (allow/deny clash on the same flow); \textit{hierarchical} (specific-scope intent overrides a group-level block via ancestor–descendant containment); \textit{synonym} (lexically distinct but semantically equivalent service names, e.g., \texttt{ssh} vs.\ \texttt{secure shell}, hiding a collision); \textit{path} (incompatible middlebox-chain requirements for the same flow); \textit{time} (overlapping temporal windows with conflicting effects on the same group); and \textit{QoS} (incompatible bandwidth directives that overcommit link capacity).

\parhead{Security Threat Framing}
Beyond expanding existing conflict detection data, we map six violation types to three security threat classes: Privilege Escalation (PE) = negation + hierarchical + synonym (broad access overrides a specific block); Isolation Bypass (IB) = path (traffic routed through untrusted segments); Denial of Service (DoS) = time + QoS (resource overcommitment). This mapping, detailed in Table~\ref{tab:mini-bench}, transforms NetInspector-Bench from a conflict detection benchmark into a security-oriented evaluation suite. Of the 437 total violations across 2,224 samples, 173 are PE, 87 are IB, and 140 are DoS.



\subsection{Evaluated Approaches}
\parhead{Baseline} For Intent Translation, we use DialogFlow~\cite{jacobs2018refining}, a commercial natural language understanding (NLU) platform that supports intent classification and entity extraction using deep learning, specifically sequence-to-sequence RNN models. For Intent Resolution, we employ Lumi~\cite{jacobs2021hey}, a rule-based system that integrates a machine learning model, specifically a Random Forest classifier, to detect conflicting or redundant intents within multi-user systems.

\parhead{LLM-Based} To answer Q2, we evaluated different prompt tuning techniques including zero-shot, few-shot w/o RAG, and few-shot w/ RAG. Each prompt defines the LLM's role as a network assistant and supplies the relevant network context as structured JSON, alongside task-specific instructions. The detailed structure of each prompt template is presented in Table~\ref{tab:prompt_templates}. We also consider fine-tuning, which is the state-of-the-art continuous learning approach. For fine-tuning, we selected GPT-4o after carefully balancing performance and cost considerations. The training parameters are listed in Table~\ref{tab:fine_tuning}.

\subsection{Results}
\subsubsection{Intent Translation}
The intent translation results in Table~\ref{tab:rouge1_results} reveal a stark performance hierarchy. Zero-shot and Few-shot approaches struggle significantly, yielding low F1 scores (0.10–0.19) and poor recall (0.06–0.12), underscoring the difficulty of unguided domain-specific translation. Retrieval-Augmented Generation (RAG) marks a major inflection point, jumping to F1 scores of 0.53–0.57. High precision (0.79–0.82) confirms that dynamically retrieving relevant examples effectively grounds the model, though recall remains a bottleneck.

Crucially, only Fine-tuned GPT-4.1 achieves superior performance (F1 0.67), surpassing the traditional LSTM baseline (0.60). With high precision (0.89) and balanced recall (0.53), it demonstrates that while RAG offers improvement, domain-specific weight adaptation is essential for practical utility. These findings indicate that general-purpose LLMs require substantial customization to match specialized non-LLM baselines in Intent-Based Networking.

\begin{table}[t]
\centering
\caption{Intent Translation Results: Baseline uses an encoder-decoder RNN with LSTM and word-level tokenization.}
\normalsize  
\setlength{\tabcolsep}{6pt}  
\begin{tabular}{l@{\hskip 6pt}c@{\hskip 6pt}c@{\hskip 6pt}c@{\hskip 6pt}c}
\toprule
\textbf{Method} & \textbf{Model} & \multicolumn{3}{c}{\textbf{ROUGE-1}} \\
\cmidrule(lr){3-5}
& & \textbf{Precision} & \textbf{Recall} & \textbf{F1} \\
\midrule
 & gpt-4o-mini & 0.35 & 0.06 & 0.10 \\
Zero-Shot & gpt-4o & 0.41 & 0.08 & 0.13 \\
 & claude-3.7 & 0.38 & 0.07 & 0.12 \\
\midrule
 & gpt-4o-mini & 0.42 & 0.09 & 0.15 \\
Few-Shot & gpt-4o & 0.47 & 0.12 & 0.19 \\
 & claude-3.7 & 0.45 & 0.11 & 0.18 \\
\midrule
 & gpt-4o-mini & 0.79 & 0.40 & 0.53 \\
RAG & gpt-4o & 0.82 & 0.44 & 0.57 \\
 & claude-3.7 & 0.81 & 0.43 & 0.56 \\
\midrule
Fine-tuning & \makecell[c]{gpt-4.1-\\2025-04-14} & 0.96 & 0.78 & 0.86 \\
\midrule
Baseline\cite{jacobs2018refining} & RNN & 0.85 & 0.47 & 0.60 \\
\bottomrule
\end{tabular}
\label{tab:rouge1_results}
\end{table}

\begin{tcolorbox}[
    colback=blue!3!white,
    colframe=blue!50!white,
    coltitle=white,
    title=\textbf{Finding I.},
    fonttitle=\bfseries\small,   
    boxrule=0.3pt,               
    arc=0.6pt,
    left=1pt, right=1pt,         
    top=1pt, bottom=1pt,         
    boxsep=0.8pt                 
]
\small   
LLMs demonstrate strong performance after fine-tuning on tasks grounded in natural language understanding, such as intent translation. This aligns with their pretraining on large-scale textual corpora and reinforces their strength in semantic parsing and generation.
\end{tcolorbox}

\subsubsection{Intent Resolution}

\begin{table}[ht]
\centering
\caption{Intent Resolution Results: In Baseline, RF represents Random Forest, LR represents Linear Regression, SVM represents State Vector Machine. Baseline models rely on numerical features manually defined and extracted from the input intent pairs.}
\normalsize
\setlength{\tabcolsep}{3pt} 
\renewcommand{\arraystretch}{1.1}
\resizebox{\columnwidth}{!}{ 
\begin{tabular}{c c c c c c c}
\toprule
\textbf{Method} & \textbf{Model} & \textbf{FPR} & \textbf{TPR} & \textbf{FNR} & \textbf{TNR} & \textbf{F1} \\ 
\midrule
\multirow{3}{*}{Zero-Shot} 
& Llama 3.2-3B & 0.47 & 0.55 & 0.45 & 0.55 & 0.54 \\
& gpt-4.1-nano & 0.29 & 0.48 & 0.52 & 0.71 & 0.54 \\
& gpt-4o-mini  & 0.47 & 0.68 & 0.32 & 0.53 & 0.63 \\
\midrule
\multirow{3}{*}{Few-Shot} 
& Llama 3.2-3B & 0.46 & 0.54 & 0.46 & 0.54 & 0.53 \\
& gpt-4.1-nano & 0.14 & 0.47 & 0.53 & 0.86 & 0.59 \\
& gpt-4o-mini  & 0.12 & 0.47 & 0.53 & 0.88 & 0.59 \\
\midrule
\multirow{3}{*}{RAG} 
& Llama 3.2-3B & 0.41 & 0.48 & 0.52 & 0.59 & 0.51 \\
& gpt-4.1-nano & 0.20 & 0.46 & 0.54 & 0.80 & 0.55 \\
& gpt-4o-mini  & 0.06 & 0.42 & 0.58 & 0.94 & 0.57 \\
\midrule
Fine-tuning & \makecell[c]{gpt-4.1\\-2025-04-14} & 0.05 & 0.43 & 0.57 & 0.95 & 0.58 \\
\midrule
\multirow{3}{*}{Baseline\cite{jacobs2021hey}} 
& RF & 0.16 & 0.93 & 0.06 & 0.83 & 0.88 \\
& LR & 0.25 & 0.89 & 0.10 & 0.74 & 0.82 \\
& SVM & 0.80 & 1.00 & 0.00 & 0.19 & 0.70 \\
\bottomrule
\end{tabular}
}
\label{tab:overall_performance}
\end{table}

Table~\ref{tab:overall_performance} presents the performance of LLM-based approaches compared to traditional baselines on the security invariant enforcement task, revealing a distinct trade-off between precision and recall across prompting strategies. Fine-tuning, few-shot prompting, and RAG exhibit a strong conservative bias. While these approaches achieve high precision and minimal false positive rates, indicating high trustworthiness in their positive predictions, they consistently suffer from low recall. This suggests that incorporating domain context, whether through weight updates or in-context examples, successfully suppresses false alarms but constrains the models from identifying more complex or subtle security violations. Conversely, zero-shot prompting demonstrates the inverse behavior; lacking these constraints, it achieves higher recall but succumbs to poor precision and a high false positive rate. Ultimately, all LLM-based methods are significantly outperformed by the traditional Random Forest baseline, which leverages labeled data to learn structured patterns, achieving a superior balance of accuracy and recall that the generative models fail to match.

We attribute the underperformance of LLMs to a deficit in spatial and topological reasoning within the network context. For example, a security violation often arises when two intents compete for the same critical link in their routing paths. To detect this, a model must deduce the specific path for each intent based on the source, destination, and network topology. This requires a structured, multi-step reasoning process that standard LLMs struggle to execute spontaneously. Consequently, while fine-tuning and prompting strategies can improve precision, they do not fundamentally resolve the lack of deep topological understanding required for robust security invariant enforcement. We provide a more detailed analysis of the scenarios where LLMs fail in the following section.

\begin{tcolorbox}[
    colback=blue!3!white,
    colframe=blue!50!white,
    coltitle=white,
    title=\textbf{Finding II.},
    fonttitle=\bfseries\small,   
    boxrule=0.3pt,               
    arc=0.6pt,
    left=1pt, right=1pt,         
    top=1pt, bottom=1pt,         
    boxsep=0.8pt                 
]
\small   
Even fine-tuned models, while more consistent, struggle to detect indirect constraint violations, revealing persistent reasoning bottlenecks across prompting and training strategies.
\end{tcolorbox}

\subsection{Failed Case Analysis}
We manually examined the cases where LLMs failed to detect security invariant violations, and summarized the recurring scenarios where LLMs tend to miss. We believe these failure modes are beneficial to the community and should be considered by anyone integrating LLMs into security-critical networking tasks.

\parhead{Availability Invariant Violations (DoS Risk)}
LLMs consistently fail to enforce Availability Invariants, treating resource constraints as abstract numbers rather than physical limitations. As illustrated in Figure~\ref{fig:example-failed-cases}(a), Intent A mandates a minimum guarantee (40 Mbps) while Intent B sets a maximum cap (60 Mbps) for the same group. A naive LLM deems these compatible ($40 \le 60$), blindly approving the configuration. However, from a security perspective, this ignores the \textit{physical link capacity}. If the underlying link supports only 50 Mbps, the LLM has inadvertently authorized a configuration that guarantees resource exhaustion, effectively creating a precondition for Denial-of-Service (DoS). This state amnesia leaves the network control plane vulnerable to availability attacks where authorized policies cannibalize critical resources.

\parhead{Privilege Escalation via Scope Blindness}
Figure~\ref{fig:example-failed-cases}(b) demonstrates a critical Access Control Bypass vulnerability rooted in the LLM's inability to resolve topological group membership. In this scenario, a group-level policy (e.g., Block 'dorms') dictates a security boundary, while a specific flow intent (e.g., Allow endpoint X) requests access. The LLM, lacking access to the runtime state of endpoint X, assumes no explicit overlap equals no conflict. This violates the fail-safe defaults principle. By defaulting to a non-conflicting judgment without verifying membership, the LLM permits the specific flow to bypass the group-level restriction, resulting in unauthorized privilege escalation or isolation breach.

\parhead{Enforcement Granularity Evasion}
LLMs struggle to maintain policy integrity across different layers of abstraction (Entity vs. Group), creating Inconsistent Enforcement vulnerabilities. As shown in Figure~\ref{fig:example-failed-cases}(c), an attacker might exploit the ambiguity between a specific IP quota and a broad group quota. Without explicit grounding, the LLM treats these as independent variables rather than hierarchical constraints. This failure allows conflicting security postures to coexist on the same traffic flow, depending on which rule takes precedence in the enforcement engine. Such non-deterministic behavior degrades the network's security posture, making it impossible to formally verify that isolation or consumption limits are being enforced reliably.


Analysis of recurring failure modes reveals that LLMs lack a structured internal understanding of network topology, intent scope, and policy semantics. This limitation is particularly pronounced when correct adjudication requires reasoning about physical feasibility, resource allocation, or complex policy interactions rather than surface-level syntax. 
Absent explicit modeling of network relationships and constraints result in failures that require topological awareness and reasoning over implicit policy dynamics.

\begin{tcolorbox}[
    colback=blue!3!white,
    colframe=blue!50!white,
    coltitle=white,
    title=\textbf{Finding III.},
    fonttitle=\bfseries\small,   
    boxrule=0.3pt,               
    arc=0.6pt,
    left=1pt, right=1pt,         
    top=1pt, bottom=1pt,         
    boxsep=0.8pt                 
]
\small   
LLMs lack structured internal representations of network topology and state, which limits their ability to reason about feasibility, policy enforcement, or intent interactions that go beyond surface-level text patterns.
\end{tcolorbox}

\section{NetInspector}


\begin{figure*}[ht]
	\centering
	\includegraphics[width=0.9\textwidth]{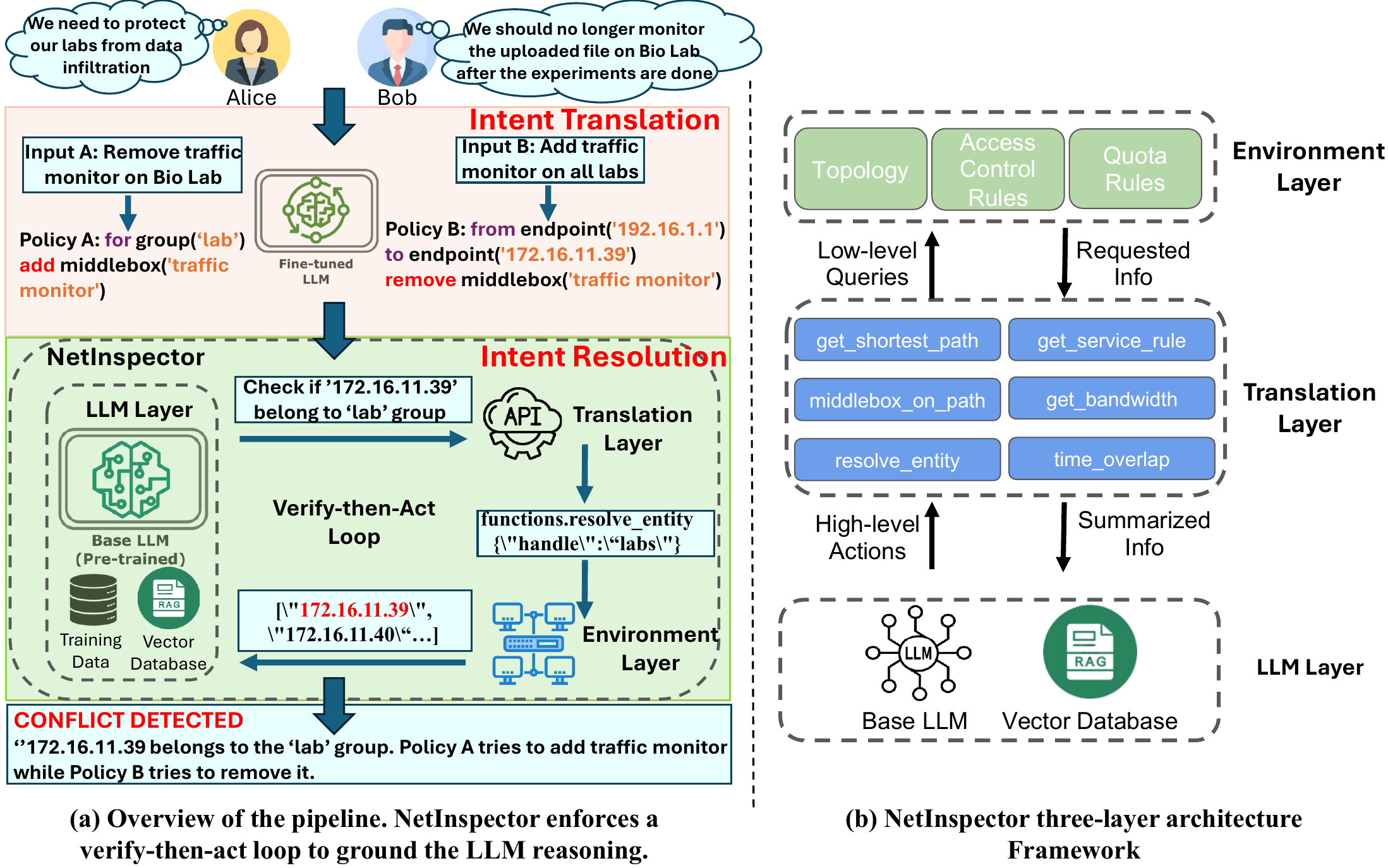}
	\caption{Overview of the system and NetInspector Framework.}\label{fig:pipeline-new}
\end{figure*}

\subsection{System Overview}
Figure~\ref{fig:pipeline-new}(a) illustrates the overall pipeline of our system, which functions as a secure gateway for transforming unreliable user intents into verified network configurations. The workflow begins with the \textit{Intent Translation} phase, where a fine-tuned LLM converts natural language requests into structured network policies. In our implementation, we utilize the Lumi~\cite{Wool2011RuleComplexity} schema as the intermediate representation due to its deterministic grammar and well-established ecosystem. However, unlike traditional automation pipelines that might blindly deploy this output, we treat the translated policy as an untrusted object that must be vetted against the network's security invariants.

Consequently, the core of our architecture is NetInspector, which enforces a verify-then-act loop before any policy reaches the network. NetInspector is organized into three layers, each with a distinct role: the Environment Layer maintains a live, queryable store of network state; the Translation Layer exposes a typed tool set through which the agent retrieves topology, group membership, and resource constraints deterministically; and the LLM Layer drives the verification loop as a general-purpose reasoning engine. A policy is approved for deployment only after the agent has grounded its decision in observations retrieved from the Environment Layer, explicitly confirming that no security invariants are violated, such as isolation boundaries or availability constraints.

\subsection{Design Rationale}
Section~\ref{sec:measurement} identified a consistent failure root cause: LLMs lack persistent grounding in network state, causing them to approximate and frequently miss violations that require traversing topology, resolving group containment, or comparing physical resource limits. Three design requirements follow directly. First, every policy decision must be conditioned on live, verifiable network facts rather than the LLM's parametric memory; this motivates the Environment Layer as an authoritative, queryable store of topology, access-control rules, and bandwidth quotas. Second, group containment and entity scope must be resolved deterministically rather than inferred from surface text—the root cause of Privilege Escalation, Isolation Bypass, and DoS misses in Section~\ref{sec:measurement}; this motivates the Translation Layer's typed tool set, which forces the agent to retrieve ground-truth membership, path data, and physical capacity before issuing any verdict. 
Thirdly, naively checking $P_\text{new}$ against every invariant in $\mathcal{L}$ incurs $N$ LLM calls per intent; since policies with disjoint entity sets cannot conflict under any topology, an entity-overlap pre-filter can safely prune the majority of pairs before any LLM invocation, keeping verification latency practical as $\mathcal{L}$ grows. 
Finally, we enforce the verify-then-act protocol by including a Verdict Node that ensures the LLM can only emit a terminal verdict after at least one grounded observation, preventing hallucinated or malformed output from reaching the network.

\subsection{NetInspector}
\label{sec:netinspector}

Figure~\ref{fig:pipeline-new}(b) details the architecture of NetInspector, a three-layer agentic framework designed to secure the interface between probabilistic planning and deterministic network enforcement. The design strictly enforces a separation of concerns principle where the Environment Layer acts as the immutable ground truth, the Translation Layer functions as a secure tool-use interface, and the LLM Layer serves as the symbolic reasoning engine. This layered architecture is critical for security assurance because it effectively creates a verify-then-act control loop. By placing a deterministic translation mechanism between the open-ended reasoning of the LLM and the concrete state of the network, we eliminate the model's ability to guess topological facts. Instead, the framework forces the agent to plan a query, execute a validated tool, and receive a confirmed observation before it can issue any policy verdict. This ensures that every decision is causally linked to verifiable network evidence rather than probabilistic inference.

\begin{table*}[t]
    \centering
    \caption{\textbf{NetInspector System Prompt Design.} The prompt enforces a structured verify-then-act protocol, ensuring semantic parsing precedes topological data retrieval.}
    \label{tab:prompt}
    \small
    \renewcommand{\arraystretch}{1.1}
    \begin{tabularx}{\textwidth}{l X}
        \toprule
        \textbf{Component} & \textbf{Key Instructions \& Content} \\
        \midrule
        \textbf{Role \& Objective} & 
        Act as \textbf{NetInspector}, an expert in network policy conflict detection. Your task is to decide whether two high-level intents conflict. Always understand each intent precisely (Subject, Action, Resource, Scope) before deciding. \\
        \midrule
        
        \textbf{Phase 1: Semantic} & 
        \textbf{Step 0 (Parse):} Extract \textit{Subject, Action, Resource, Scope} in plain language. \newline
        \textbf{Step 1 (Logic):} Compare parsed records using text only. Declare \texttt{conflict=true} if:
        \begin{itemize}
            \item Same subject receives incompatible directives (e.g., Add vs Remove).
            \item Numeric limits clash (e.g., \texttt{quota('any')} overrides specific directions).
            \item Time windows overlap while prescribing contradictory behaviors.
        \end{itemize} \\
        \midrule
        
        \textbf{Phase 2: Topology} & 
        \textbf{Step 2 (Network-Aware Check):} If text analysis is inconclusive, use deterministic tools:
        \begin{itemize}
            \item \textbf{Shared Path:} Use \texttt{get\_shortest\_path} to check if separate flows share compromised links.
            \item \textbf{Scope Expansion:} Use \texttt{resolve\_entity} to check if a group label subsumes a specific host.
            \item \textbf{Capacity:} Use \texttt{get\_effective\_bandwidth}. If \texttt{min + max} $\ge$ \texttt{capacity}, declare conflict.
        \end{itemize} \\
        \midrule
        
        \textbf{Few-Shot Examples} & 
        \textbf{Scope Hierarchy:} Specific upload quota vs. Global `any` quota $\rightarrow$ \textit{Verdict: CONFLICT}. \newline
        \textbf{Shared Path:} Add IPS on Flow X vs. Remove IPS on Flow Y (sharing links) $\rightarrow$ \textit{Verdict: CONFLICT}. \newline
        \textbf{Resource Exhaustion:} Min 40Mbps + Max 60Mbps vs. 100Mbps Capacity $\rightarrow$ \textit{Verdict: CONFLICT}. \\
        \bottomrule
    \end{tabularx}
\end{table*}

\parhead{Environment Layer}
The Environment Layer captures all runtime state relevant to policy generation: the physical topology, service-level access-control rules, and bandwidth-quota policies. Unlike the parametric memory of an LLM, which is static and prone to obsolescence, this layer provides a live, queryable view $\mathcal{E}$ over the network state $\sigma$ and the standing policy library $\mathcal{L}$ defined in \S\ref{sec:threat-model}. We model the physical topology as a directed graph in NetworkX\cite{NetworkX}, where nodes represent entities such as switches, hosts, and middleboxes, each annotated with immutable security roles and logical memberships. Edges within this graph store physical properties like capacity and latency, enabling the system to definitively validate path-dependent constraints that purely semantic parsers inevitably miss, such as isolation boundaries or waypoint enforcement. Specifically, we represent every switch, host, and middle-box as a node annotated with its IP prefix, functional role (e.g., \textit{edge-fw}, \textit{core-sw}), and logical group membership (\textit{students}, \textit{labs}, \textit{dorms}).  Each vertex, therefore, records an immutable identifier, role tag, and group set, while each edge stores link capacity and latency as weights.  This graph supports path-based queries such as reachability and middle-box placement without embedding device-specific semantics.

Beyond topology, security policies are maintained as a priority-ordered JSON list whose entries have the schema ⟨\textit{src}, \textit{dst}, \textit{service}, \textit{action}, \textit{enforced\_at}, \textit{start}, \textit{end}⟩.  A rule binds a logical source (group label or host IP) to a destination scope (\textit{\*} for any or a group label), names the application service, specifies the decision, and locates the enforcement point (e.g., border or core firewall).  Time windows are optional and recorded in local wall-clock time.  Bandwidth controls follow an analogous tuple format ⟨\textit{endpoint}, \textit{traffic\_type}, \textit{limit\_mbps}, \textit{start}, \textit{end}⟩. Here, the endpoint may reference a logical group or subnet, the traffic type labels the application class, and the limit is expressed in megabits per second.  Temporal scopes are likewise optional, enabling both permanent and time-bounded quotas. This structure allows the system to deterministically resolve first-match logic, which is essential for detecting policy shadowing where a broad rule might inadvertently override a specific block. Distinct from the policy list, bandwidth and usage limits are stored as tuple records, creating a dedicated store that prevents the agent from conflating independent constraints based on loose linguistic similarities. To maintain data integrity, this layer is managed by a dedicated state manager that applies updates atomically, ensuring that the agent always queries a consistent view of the network and preventing race conditions during the verification process.

We would also like to highlight that the environment layer is maintained as a live, mutable store rather than a static snapshot. To do so, a dedicated state manager mediates all updates originating from network controllers or administrative tools: new service rules are appended to the JSON policy list, bandwidth quotas are patched in place, and topology changes trigger incremental edits to the NetworkX graph. Each mutation is versioned and applied atomically, so that every query to $\mathcal{E}$ is deterministic in the observed snapshot and side-effect-free---properties on which the verification pipeline below depends.

\parhead{Translation Layer} The Translation Layer acts as the secure API gateway that mediates all interactions between the LLM Layer and the Environment Layer. As shown in Figure~\ref{fig:pipeline-new}(b), it exposes a typed tool set $\mathcal{T} = \{t_1, \dots, t_6\}$, where each $t_j : \mathcal{I}_j \to \mathcal{O}_j$ specifies an input schema $\mathcal{I}_j$ for admissible arguments and an output schema $\mathcal{O}_j$ for the canonical reply. From a security perspective, $t_j$ executes only on inputs $x \in \mathcal{I}_j$; any malformed action is rejected before reaching $\mathcal{E}$. This input sanitization, combined with the deterministic dispatch to $\mathcal{E}$, is the architectural defense that bounds what the LLM Layer can observe. The six handles instantiate the operations illustrated in Fig.~\ref{fig:pipeline-new}(b): \textit{resolve\_entity} expands logical groups into concrete host or subnet identifiers; \textit{get\_shortest\_path} and \textit{middlebox\_on\_path} query the NetworkX topology for reachability and inline devices; \textit{get\_service\_rule} returns the highest-priority access-control entry applicable to a source--destination--service triple; \textit{get\_bandwidth} retrieves any bandwidth quota governing a given endpoint and traffic class; and \textit{time\_overlap} checks whether two temporal scopes intersect. This mechanism prevents scope-related vulnerabilities by ensuring the agent reasons about the actual endpoints involved rather than just their semantic labels. Furthermore, the layer processes the often voluminous raw data from network queries, such as the output of Dijkstra's algorithm, and returns a canonical JSON reply containing both the raw result and a compact textual summary.

\parhead{LLM Layer} At the reasoning tier, we use an off-the-shelf LLM, prompted to behave as \textit{NetInspector}, as the agent $\mathcal{M}$. By stripping this layer of the burden of memorizing network state, we allow $\mathcal{M}$ to focus on symbolic planning and semantic reasoning. Concretely, $\mathcal{M}$ is a function $(P_\text{new}, h_k) \to a_{k+1}$, where the transcript $h_k = ((t^{(1)}, o^{(1)}), \dots, (t^{(k)}, o^{(k)}))$ records the sequence of tool actions and observations so far, and the action space $\mathcal{A} = \mathcal{T} \cup \{\textsc{Conflict}, \textsc{Safe}\}$ partitions every possible output into either a typed tool call or a terminal verdict. The system prompt initializes $\mathcal{M}$ with a rigorous security decision protocol. It supplies two critical ingredients: (i) domain knowledge---five conflict categories and a two-stage decision protocol, and (ii) the exact invocation syntax for the six translation-layer tools. Table~\ref{tab:prompt} shows the excerpt that is embedded in the main text. By grounding the model in this structured specification, we eliminate the need for fine-tuning and instead rely on prompt engineering plus a small in-context library of worked examples stored in the vector database. Upon receiving $P_\text{new}$, $\mathcal{M}$ first performs semantic parsing to identify obvious linguistic contradictions. If none are found, it proceeds to an Information Retrieval Plan, identifying exactly what state information is missing, such as link utilization or group membership. It then enters a tool execution loop: at each step $k$, $\mathcal{M}(P_\text{new}, h_k)$ either selects a new tool action $(t_j, x) \in \mathcal{T} \times \mathcal{I}_j$ that extends $h_k$ with a fresh $(t_j, o_j)$ observation, or emits a terminal verdict in $\{\textsc{Conflict}, \textsc{Safe}\}$. The Verdict Node enforces that only the latter case terminates the loop; any non-verdict output is rerouted through $\mathcal{T}$, so a malformed or hallucinated action never escapes. This disciplined ``Thought / Action / Observation'' loop effectively neutralizes the risk of ungrounded inference: the model serves as the logic engine that connects the dots, but the dots themselves are provided by the trusted Environment Layer. This architectural decision shifts the system's failure mode from silent policy violation, where the model confidently guesses incorrectly, to explicit uncertainty, where the model pauses to request data, significantly enhancing the system's fail-safe properties.

\parhead{NetInspector as a Verifier}
Together, the three layers compose into the NetInspector verifier $v_\text{NI}(P_\text{new}, \mathcal{L}, \sigma)$, which realizes the abstract verifier $v$ of \S\ref{sec:threat-model}. The verifier iterates $\mathcal{M}$ over each candidate invariant, routing every tool action through $\mathcal{T}$ to $\mathcal{E}$, and emits $\textsc{Conflict}$ as soon as any pairwise check returns a conflicting verdict. Every verdict is therefore conditioned on a transcript $h_k$ whose observations come deterministically from $\mathcal{E}$; soundness and completeness of $v_\text{NI}$ relative to the ideal $v$ are measured empirically in \S\ref{sec:eval}.

\parhead{Entity Pre-filter}
To scale verification against a large policy library, we observe that all six violation types require at least one shared entity as a prerequisite (negation and synonym require the same service and scope, path requires the same flow endpoints, hierarchical requires overlapping groups, and QoS and time require the same group or traffic class). Therefore, we apply an entity-overlap pre-filter before any LLM invocation: $\phi(P_\text{new}, \mathcal{L}) = \{P_i \in \mathcal{L} : \mathrm{entities}(P_\text{new}) \cap \mathrm{entities}(P_i) \neq \emptyset\}$, where $\mathrm{entities}(P)$ denotes the set of entity labels appearing in $P$; the verifier then passes only $\phi(P_\text{new}, \mathcal{L})$ to $\mathcal{M}$. In doing so, we safely pruned disjoint $P_i$ from consideration and reduce the LLM calls by around 90\%, as illustrated in Section~\ref{subsec:rq5}.

\section{Evaluation}
\label{sec:eval}
To evaluate our proposed framework, we focus on the following research questions.

\begin{itemize}
    \item \textbf{RQ1: (Effectiveness) How effectively does NetInspector detect security invariant violations compared to existing methods?}
    \item \textbf{RQ2: (Robustness) Does NetInspector maintain security coverage under linguistic distribution shifts?}
    \item \textbf{RQ3: (Component Contribution) How does each component contribute to NetInspector's security enforcement capability?}
    \item \textbf{RQ4: (Model Sensitivity) How does the choice of base LLM affect security enforcement performance?}
    \item \textbf{RQ5: (Scalability) How does NetInspector scale as the standing security invariant library grows?}
\end{itemize}

\subsection{Experiment Setup}
We implemented our framework utilizing LangGraph\cite{langgraph} for agentic workflow orchestration and ChromaDB\cite{ChromaDB} as the vector database for efficient context retrieval. To evaluate the robustness of our approach across varying model capabilities, we experimented with five different LLMs as the underlying base (gpt-4.1-mini, gpt-5-mini, gemini-2.5-flash-lite, claude-haiku-4.5, qwen3-8b). This selection comprises four state-of-the-art frontier models and one representative open-source model. We compared our method against three existing baselines: Lumi\cite{jacobs2021hey}, PGA\cite{prakash2015pga}, and a heuristic-based approach\cite{zheng2022intent}. These selected baselines represent the spectrum of prior methodologies, covering both traditional machine learning (Lumi) and non-ML-based techniques (graph-based and heuristic-based). We assess performance using three primary metrics: False Positive Rate (FPR), False Negative Rate (FNR), and F1-score. These metrics were chosen to rigorously evaluate the trade-off between the system's sensitivity to real conflicts and its ability to suppress false alarms. All experiments were conducted on a workstation running Ubuntu 24.04.3 LTS, equipped with an Intel(R) Core(TM) i7-14700F (2.10 GHz) CPU and 32 GB RAM. For evaluations involving local LLMs, we used an NVIDIA GeForce RTX 5070 Ti GPU with 16 GB VRAM.

\begin{figure*}[ht]
	\centering
	\includegraphics[width=1.0\textwidth]{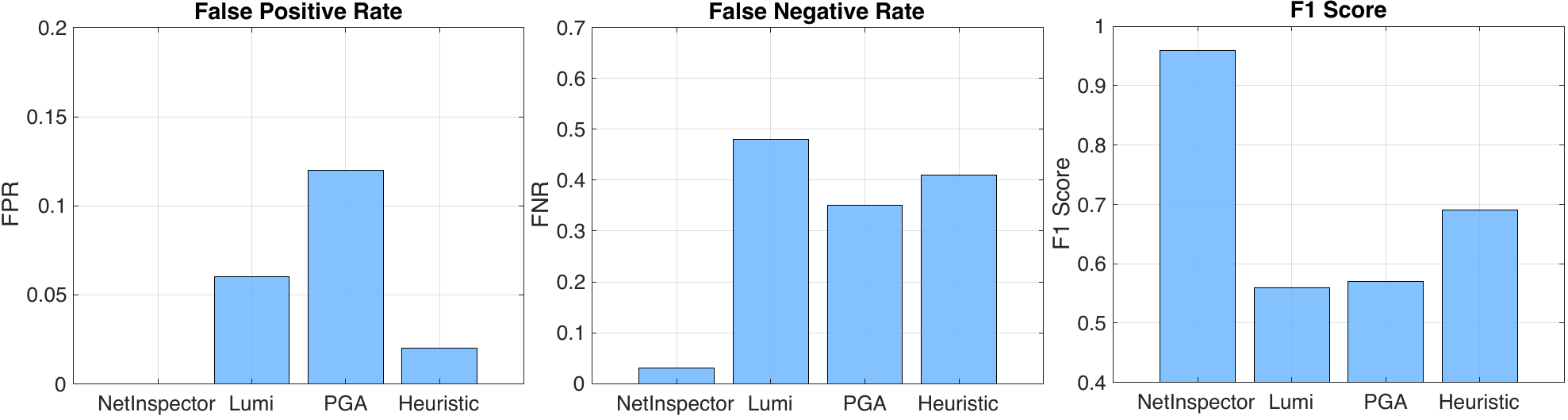}
	\caption{Security invariant enforcement performance (FNR, FPR, F1) comparing NetInspector against baselines~\cite{jacobs2021hey,prakash2015pga,zheng2022intent}. Lower FNR means fewer security-violating intents reach deployment.}\label{fig:eval-rq1}
\end{figure*}

\subsection{Effectiveness of NetInspector (RQ1)}
\label{sec:eval-rq1}

Figure~\ref{fig:eval-rq1} presents the comparative performance of NetInspector against the Lumi, PGA, and Heuristic baselines. From a security perspective, FNR is the most critical metric: each undetected violation is a security-compromising intent silently committed to the network. The results reveal a significant enforcement gap in existing methods, with baselines exhibiting FNRs between 35\% and 48\%, indicating that they fundamentally struggle to catch violations that depend on physical topology or group membership. In contrast, NetInspector reduces the FNR to approximately 3\% by grounding its reasoning in a verify-then-act loop, successfully catching the topology-dependent violations that standard classifiers consistently miss.
Furthermore, NetInspector minimizes operational alert fatigue by maintaining a near-zero false positive rate. While the Heuristic baseline achieves low noise through excessive conservatism, the ML and graph-based models suffer from significant noise due to spurious text-level correlations. Consequently, NetInspector attains an F1 score exceeding 0.95, significantly outperforming all non-LLM baselines. These results confirm that our three-layer agentic framework provides the semantic flexibility required for diverse intents while retaining the rigorous verification needed to enforce security invariants before deployment.

\begin{figure}[ht]
	\centering
	\includegraphics[width=0.45\textwidth]{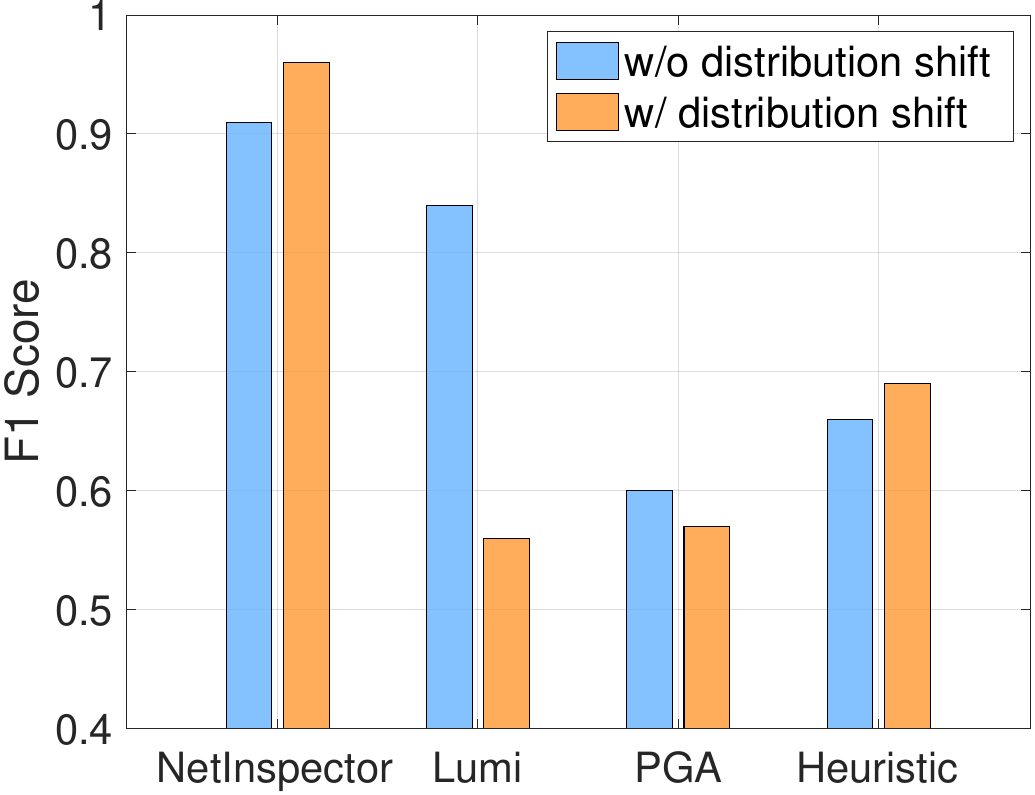}
	\caption{F1 Score under original and shifted service distributions (shift defined in Figure~\ref{fig:motivation}). Lumi's coverage collapses from 0.84 to 0.56 under shift; NetInspector is unaffected.}\label{fig:eval-rq-generalize}
\end{figure}

\subsection{Robustness under Distribution Shift (RQ2)}
\label{subsec:rq2}

Network invariants must hold across evolving operational patterns and shifting usage behaviors. To evaluate this resilience, Figure~\ref{fig:eval-rq-generalize} analyzes NetInspector's performance under a data distribution shift, representing novel traffic patterns that deviate from historical training data. As shown in Figure~\ref{fig:motivation}(a), the test set introduces significant changes in service composition, mimicking real-world scenarios where new protocols and applications are introduced.

The impact on security coverage is quantified in Figure~\ref{fig:eval-rq-generalize}(b), where traditional baselines like Lumi suffer catastrophic degradation, with F1 scores collapsing from 0.85 to 0.55. This confirms that static, feature-based classifiers function as brittle pattern-matchers that overfit to training keywords — a dangerous property for a security enforcement component. Conversely, NetInspector maintains superior robustness (F1 $>$0.95) via its three-layer agentic framework and verify-then-act architecture. Instead of memorizing surface patterns, NetInspector utilizes the Translation Layer to dynamically query live intent properties. By grounding decisions in operational network state rather than historical text statistics, the framework generalizes to unseen inputs and maintains security coverage regardless of the linguistic surface form of the intent.

\subsection{Component Contribution (RQ3)}
\label{subsec:rq3}

We investigate the contributions of individual framework components by comparing three design variants: (i) a Simple Prompt baseline; (ii) a Structured Prompt providing raw network state and Chain-of-Thought guidance; and (iii) the full NetInspector three-layer agentic framework. Results in Table~\ref{tab:eval-rq2-3} indicate that while raw context improves performance, the specific presentation of that data is paramount. The ``w/o Translation Layer'' variant still suffers from high error rates as the LLM must simultaneously process low-level network details and high-level intent logic. This cognitive overload leads to over-sensitive decisions, proving that prompt engineering alone cannot bridge the gap between raw data and intent semantics.

Conversely, the full NetInspector framework significantly reduces both false positives and negatives by utilizing the Translation Layer to decouple information retrieval from reasoning. This architecture allows the LLM to operate as a focused planner within a verify-then-act loop, querying structured properties rather than ingesting raw state. This structural decoupling is essential for security enforcement: LLMs serve as high-level decision engines supported by deterministic retrieval components, ensuring that every verdict is grounded in verifiable network facts rather than probabilistic pattern matching. Table~\ref{tab:eval-rq2-3} reports the per-sample inference time across the three design variants: while architectural complexity increases latency, the incremental cost of the full design over the Translation-Layer-only variant is modest, indicating that the LLM's internal reasoning dominates the overhead rather than our system components.

\begin{table*}[t]
\centering
\small
\renewcommand{\arraystretch}{1.2}
\caption{Performance and per-sample inference time across LLMs and design approaches. Simple means simple prompt design. w/o Translation Layer means without the translation layer. Full Design means the full three-layer framework.}
\begin{tabular}{l|cccc|cccc|cccc}
\hline
\textbf{Model} &
\multicolumn{4}{c|}{\textbf{Simple Prompt}} &
\multicolumn{4}{c|}{\textbf{w/o Translation Layer}} &
\multicolumn{4}{c}{\textbf{Full Design}} \\
&
\textbf{FN} & \textbf{FP} & \textbf{F1} & \textbf{Latency (s)} &
\textbf{FN} & \textbf{FP} & \textbf{F1} & \textbf{Latency (s)} &
\textbf{FN} & \textbf{FP} & \textbf{F1} & \textbf{Latency (s)} \\
\hline
GPT-4.1-mini
& 146 & 11  & 0.79 & 2.46
& 87  & 72  & 0.82 & 5.68
& 66  & 56  & 0.86 & 6.96 \\
GPT-5-mini
& 133 & 5   & 0.81 & 6.19
& 49  & 68  & 0.87 & 14.07
& 53  & 24  & 0.91 & 16.86 \\
Gemini-2.5-Flash-Lite
& 72  & 573 & 0.53 & 1.65
& 105 & 154 & 0.72 & 3.07
& 116 & 83  & 0.76 & 2.00 \\
Claude-Haiku-4.5
& 172 & 57  & 0.70 & 3.87
& 23  & 144 & 0.83 & 5.05
& 19  & 29  & 0.95 & 7.97 \\
Qwen3-8B
& 131 & 83  & 0.74 & 4.66
& 76  & 135 & 0.77 & 5.34
& 84  & 80  & 0.81 & 10.52 \\
\hline
\end{tabular}
\label{tab:eval-rq2-3}
\end{table*}

\subsection{Impact of Different Base Models (RQ4)}
\label{subsec:rq4}

We further analyze how different LLM backbones perform to isolate the impact of model capacity from our architectural design. As shown in Table~\ref{tab:eval-rq2-3}, several consistent trends emerge across the evaluated models. While more capable models generally achieve more stable performance and exhibit lower sensitivity to contextual noise, they still experience significant degradation when operating under the Simple Prompt and raw-state settings. This demonstrates that model strength alone cannot overcome the inherent challenges of reasoning over ungrounded data or unformatted network state.
However, under the full NetInspector design, performance gaps between models narrow significantly as all backbones benefit from the structural guidance provided by our three-layer agentic framework. By confining the LLM to high-level planning within a verify-then-act loop, the system effectively stabilizes security enforcement across different backbones. These results suggest that while model choice influences baseline accuracy, a well-designed system architecture is the dominant factor in reliably catching security invariant violations — regardless of which foundation model is used.

\begin{table}[t]
\centering
\small
\caption{N-policy verification latency (s) vs.\ library size.}
\label{tab:rq5-latency}
\setlength{\tabcolsep}{4pt}
\begin{tabular}{lcccc}
\toprule
\textbf{Model} & \textbf{N=10} & \textbf{N=20} & \textbf{N=50} & \textbf{N=100} \\
\midrule
Gemini-2.5-Flash   &  2.8  &  6.1  &  9.3  & 22.1  \\
Claude-Haiku-4.5   &  6.0  & 10.7  & 24.5  & 48.7  \\
GPT-4.1-mini       &  6.7  & 12.0  & 26.8  & 50.4  \\
GPT-5-mini & 18.9 & 34.8 & 56.8  & 98.3  \\
Qwen3-8B       & 10.7  & 21.5  & 53.7  & 107.3 \\
\bottomrule
\end{tabular}
\end{table}

\begin{table}[t]
\centering
\small
\caption{Entity pre-filter effectiveness (campus dataset, 1{,}000 tests). ``Checks w/ filter’’ = avg.\ LLM calls per intent.}
\label{tab:rq5-prefilter}
\begin{tabular}{cccc}
\toprule
\textbf{N} & \textbf{Checks (no filter)} & \textbf{Checks (w/ filter)} & \textbf{Reduction} \\
\midrule
10  & 10  & 1.1 & 89.0\% \\
20  & 20  & 2.1 & 89.5\% \\
50  & 50  & 5.0 & 90.0\% \\
100 & 100 & 9.8 & 90.2\% \\
\bottomrule
\end{tabular}
\end{table}

\begin{figure}[ht]
	\centering
	\includegraphics[width=0.5\textwidth]{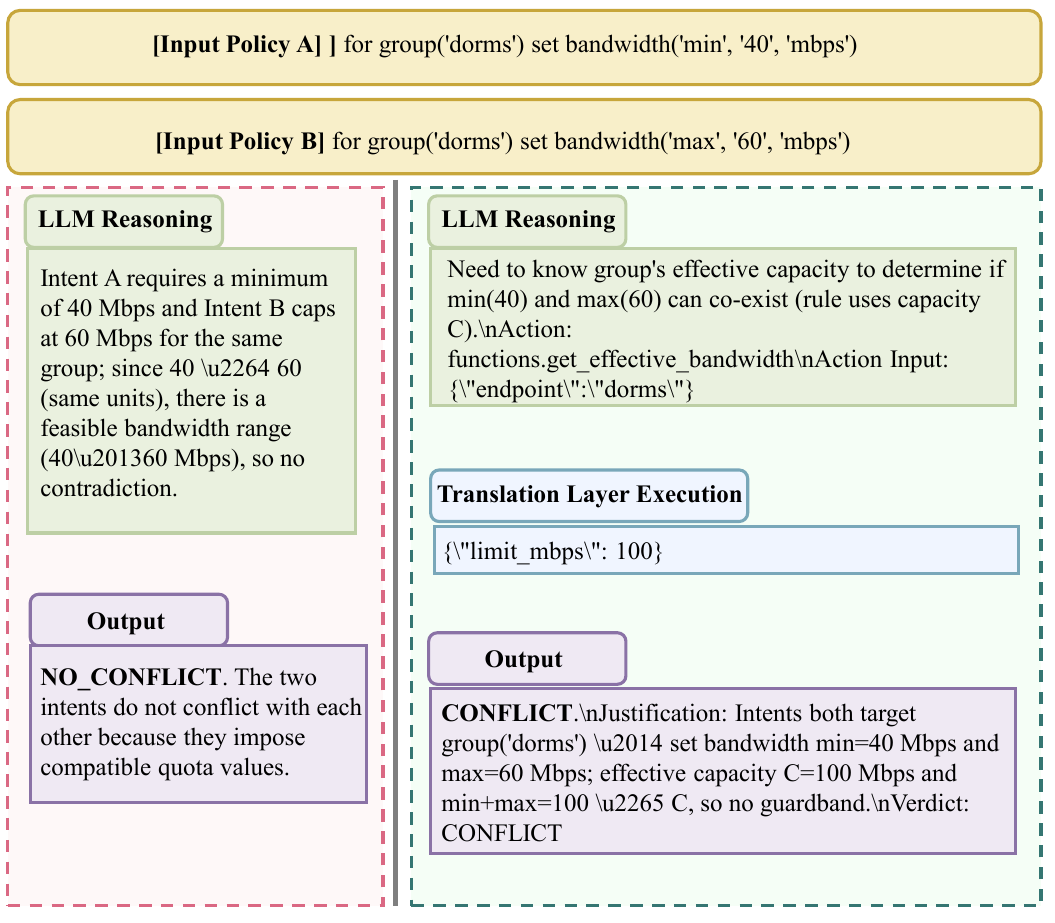}
	\caption{Summarized example of Denial of Service. The left shows the execution of LLM reasoning without NetInspector, while the right showcases how LLM interacts with NetInspector and mitigates this threat through capacity-aware reasoning.}\label{fig:case-study-resource-competition}
\end{figure}

\subsection{Scalability under N-Policy Libraries (RQ5)}
\label{subsec:rq5}

We evaluate NetInspector’s scalability when checking a new intent against a library of $N$ standing security invariants—the operational mode faced by organizations as their policy sets grow. Two properties are essential for practical deployment: (1) the entity pre-filter introduced in \S\ref{sec:netinspector} must substantially reduce the number of pairwise LLM calls, and (2) end-to-end latency must remain acceptable as $N$ grows.

\parhead{Pre-filter effectiveness}
Table~\ref{tab:rq5-prefilter} shows that the entity pre-filter passes only ${\approx}10\%$ of library pairs for LLM evaluation across all library sizes. A new intent typically references specific network entities (hosts, groups, prefixes) that overlap with only a small fraction of standing invariants, yielding a consistent $10{\times}$ reduction in LLM invocations. This efficiency is orthogonal to library size: as $N$ grows, actual LLM calls grow at ${\approx}10\%$ of $N$, not at $N$ itself.

\parhead{Latency vs.\ library size}
Table~\ref{tab:rq5-latency} reports end-to-end verification latency for five LLMs as the library size $N$ grows from 10 to 100. For context, without pre-filtering, total latency would scale as per-pair latency $\times N$ (from Table~\ref{tab:eval-rq2-3}): e.g., 200\,s for Gemini-2.5-Flash, 797\,s for Claude-Haiku-4.5, at $N{=}100$. The pre-filter reduces actual latency to $8$--$10{\times}$ below that unfiltered cost across all models. Even at $N{=}100$, the fastest model (Gemini-2.5-Flash) completes verification in 22\,s, and all measured models remain within practical bounds for control-plane operations.

\begin{figure}[ht]
	\centering
	\includegraphics[width=0.5\textwidth]{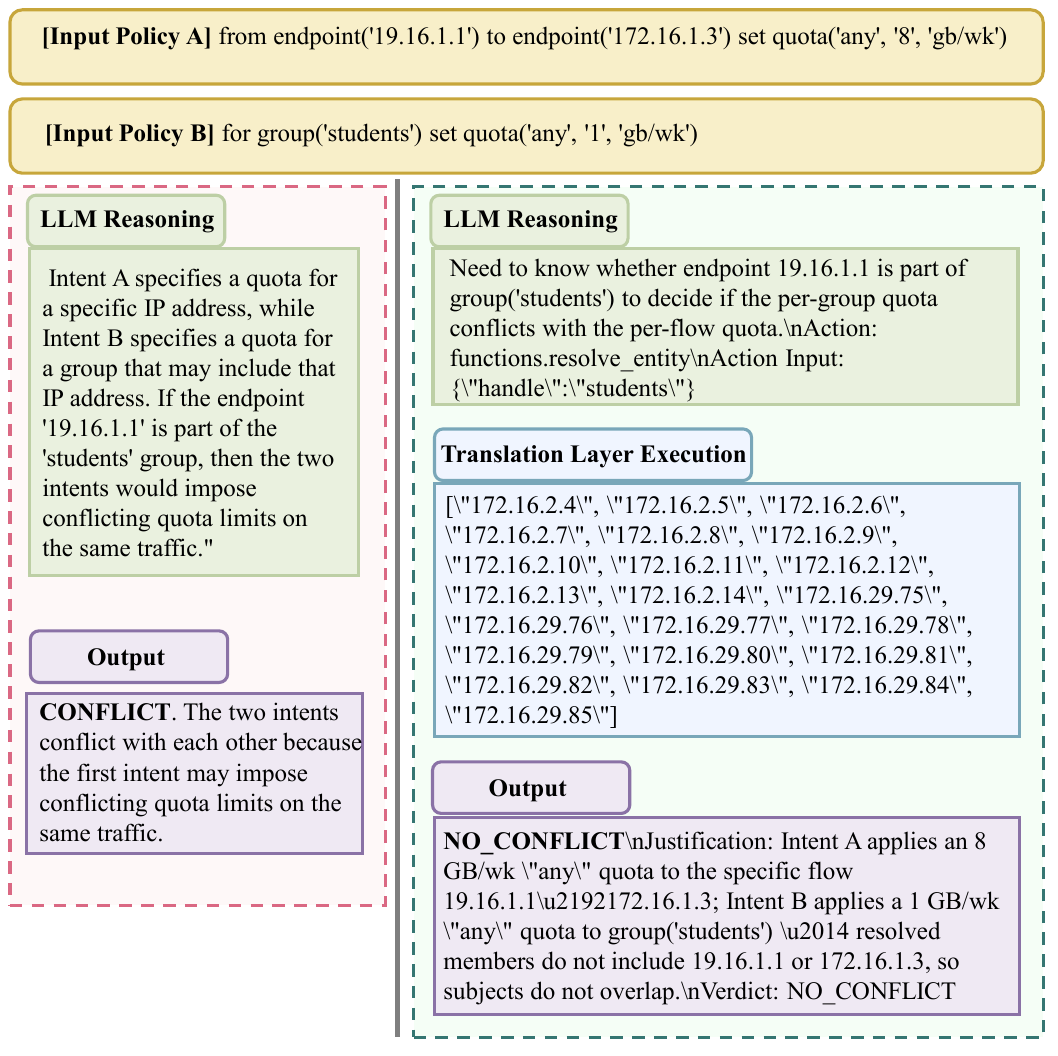}
	\caption{Summarized example of Privilege Escalation via scope disambiguation. The left shows the execution of LLM reasoning without NetInspector, while the right showcases the mitigation of LLM interacting with NetInspector.}\label{fig:case-study-resolve-entity}
\end{figure}

\subsection{Illustrative Case Study}\label{subsec:case-study}
To concretely demonstrate how LLMs interact with NetInspector to enforce security invariants, we present the following illustrative case studies, showing how each of the three threat classes is either missed or caught depending on whether NetInspector is in the loop.

\begin{figure}[ht]
	\centering
	\includegraphics[width=0.5\textwidth]{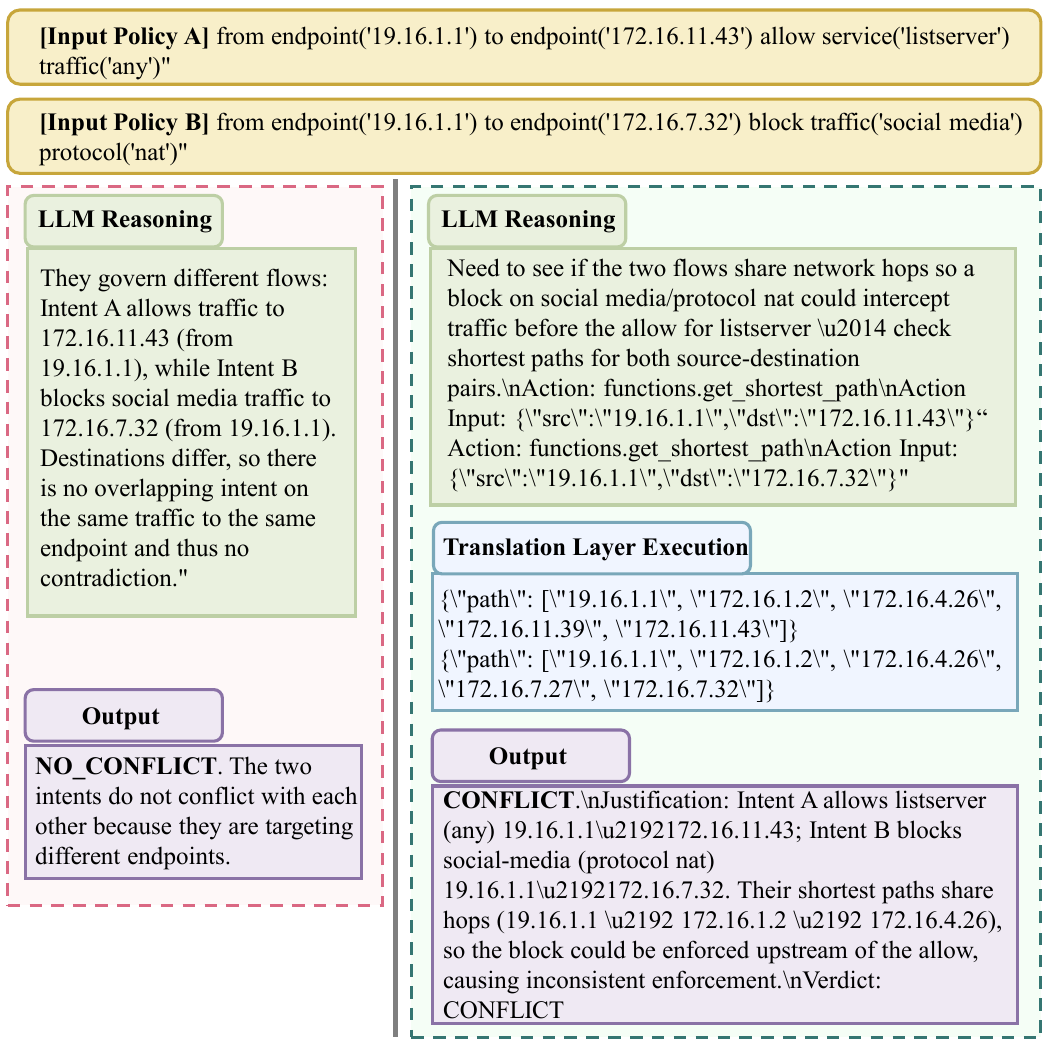}
	\caption{Summarized example of Isolation Bypass. The left shows the execution of LLM reasoning without NetInspector, while the right showcases how LLM interacts with NetInspector and mitigates this threat through topology-aware reasoning.}\label{fig:case-study-middlebox-negation}
\end{figure}

\parhead{Case Study 1: Preventing DoS via Capacity-Aware Reasoning}
Figure~\ref{fig:case-study-resource-competition} illustrates a critical defense against unintentional Denial of Service (DoS), where abstractly valid policies threaten network stability due to resource exhaustion. In this scenario, two intents target the \texttt{dorms} group: one demanding a minimum bandwidth guarantee, and another setting a maximum usage cap. A naive LLM, treating this purely as a symbolic logic problem, incorrectly concludes \texttt{NO\_CONFLICT} because the numeric bounds form a valid mathematical range. However, this failure to account for physical constraints creates a high risk of oversubscription, where the network commits to guarantees it cannot physically support.
NetInspector neutralizes this threat by enforcing state-grounded reasoning. Instead of making a premature judgment based on incomplete assumptions, the agent recognizes that validity depends on the group's effective capacity. It invokes the Translation Layer to retrieve the concrete bandwidth limit from the network state. Armed with this ground truth, the LLM determines that the combined constraints exceed available resources and correctly identifies the conflict. This structured verify-then-act workflow effectively bridges the gap between high-level intent and physical reality, preventing the cumulative resource allocation errors that lead to service degradation. 



\parhead{Case Study 2: Mitigating Privilege Escalation via Scope Disambiguation}
Figure~\ref{fig:case-study-resolve-entity} addresses the fundamental vulnerability behind unintentional Privilege Escalation: the failure to accurately resolve hierarchical set containment relationships. In this scenario, the system must arbitrate between a flow-level quota (specific endpoint-to-endpoint) and a group-level quota. A naive LLM, lacking access to the directory state, cannot verify if the specific endpoint belongs to the target group. Consequently, it resorts to conservative speculation, assuming a potential overlap and flagging a false conflict. While this specific instance results in a false alarm, this same blindness is what allows broad permissions to inadvertently override specific restrictions in adversarial contexts.
NetInspector prevents this boundary confusion through explicit entity resolution. Instead of guessing, the agent invokes the Translation Layer to query the group's concrete membership. Upon confirming that the endpoint is not a member of the group, the system correctly concludes that the scopes are disjoint. By replacing probabilistic assumptions with grounded evidence, the framework ensures that policies are applied strictly to their intended targets, enforcing the precise logical boundaries necessary to prevent privilege escalation. Figure~\ref{fig:case-study-muliti-step} presents another more complicated example of Privilege Escalation where NetInspector performs multi-step reasoning to identify the security violation.



\parhead{Case Study 3: Preventing Isolation Bypass via Topology-Aware Reasoning}
Figure~\ref{fig:case-study-middlebox-negation} demonstrates the necessity of topological grounding to prevent Isolation Bypass, where ignorance of the physical network allows traffic to violate segmentation boundaries. In this example, the system evaluates two intents governing different source–destination pairs. A naive LLM, restricted to an endpoint-centric view, incorrectly concludes \texttt{NO\_CONFLICT} because the logical endpoints appear disjoint. This flat reasoning fails to account for the physical reality that logically distinct flows may traverse shared infrastructure.
NetInspector overcomes this blindness by enforcing path-level verification. Recognizing that logical isolation does not guarantee physical separation, the agent queries the Translation Layer to retrieve the concrete forwarding paths for both flows. The returned data reveals that the flows converge on a shared intermediate hop, meaning a restrictive policy applied to one flow would inadvertently intercept the other. By exposing these hidden topological dependencies, the framework ensures that the LLM detects the violation, preventing the very class of routing errors that lead to traffic leakage and isolation bypass.

\section{Limitations and Future Directions}

\parhead{Adversarial threat model} Our threat model targets unintentional misconfigurations by benign insiders; adversarial prompt injection and intentionally crafted unsafe intents are out of scope. NetInspector's architecture provides partial defenses---a Verdict Node enforces a strict LangGraph state machine, every approved verdict must be backed by Environment Layer observations, and the Translation Layer validates tool arguments against typed schemas before execution. A deployment exposed to adversarial users would additionally require guardrail models (e.g., LlamaGuard) and standard hardening such as RBAC and rate limiting. Securing IBN against adversarial inputs is complementary to, not subsumed by, the verification NetInspector provides.

\parhead{Coverage of violation classes and toolset} The three representative violation classes used throughout this paper and the six tools in our current implementation map onto the identity, topology, and capacity primitives that underlie network policy. We do not formally prove this taxonomy is exhaustive; production environments may surface violation classes outside our current coverage, such as service-chaining conflicts, multi-tenant policy layering, or rule-precedence interactions. The architecture is modular---adding a new tool requires only a JSON schema update and a small set of in-context examples---but identifying the relevant networking primitive for each new class still requires expert judgment.

\parhead{Synthetic benchmark} NetInspector-Bench is constructed by synthesizing intents on top of three TopologyZoo graphs; we do not validate NetInspector on production IBN deployments with evolving state, complex rule-precedence stacks, or service chaining. Real environments are likely to surface failure modes not represented in our benchmark, and the FNRs we report should be treated as a lower bound on what production deployments will encounter. Closing this gap requires partnerships with operators willing to share anonymized policy logs.


\parhead{Scalability at very large libraries} Current evaluation characterizes NetInspector's latency up to libraries of $N{=}100$ standing invariants. Production deployments may carry thousands. The entity pre-filter scales cheaply (its pass rate is independent of $N$), but residual LLM-call latency continues to grow with $N$, and at very large scales the bottleneck may shift to retrieval-side I/O against the Environment Layer. One further improvement to the scaling problem is to compare input intent with existing security policies in parallel.


\parhead{Future Directions}
Beyond closing the limitations above, several directions extend NetInspector's reach. First, intents that are individually safe against each invariant but collectively unsafe (e.g., cumulative bandwidth across approved reservations) require cross-invariant aggregation outside our current pairwise scope; the Environment Layer already tracks per-link capacity, so extending it with a running admission ledger turns each capacity check into a query against accumulated reservations, capturing multi-intent resource exhaustion without new architectural components. Second, the LLM's generative capabilities could be leveraged for counterfactual reasoning---analyzing why a violation occurred and proposing safe alternatives: for a bandwidth conflict the agent could propose reducing the requested rate to residual capacity; for a group violation it could rewrite scope to exclude the protected boundary, turning NetInspector from a binary verdict oracle into a repair engine. Third, the IBN lifecycle includes \textit{Intent Activation} and \textit{Intent Assurance} stages~\cite{leivadeas2022survey} beyond translation and resolution, which the verify-then-act pattern can naturally extend to. Finally, compiling the agent's verified logic into programmable data plane primitives~\cite{zhou2023efficient, yan2024brain} would enable line-rate enforcement of validated invariants. A natural scalability improvement is to parallelize the post-filter pairwise checks: since each is independent, latency drops from $O(N \cdot t)$ to $O(t)$, bounded by the slowest single check.

\section{Conclusion}
In this work, we seek to answer if LLM can be reliably used for Intent-Based Networking policy generation, which includes both translation of high-level intent and detect any violations to existing security invariants. Through our measurement study, we discovered that ungrounded models exhibit high FNR as security invariant enforcers, allowing unsecured intents to reach deployment undetected. To close this gap, we introduced NetInspector, a three-layer agentic framework that acts as a security enforcement layer between operator intent and policy deployment. By decoupling high-level planning from deterministic state retrieval, NetInspector enforces a strict verify-then-act protocol that grounds every policy decision in verifiable network facts. Our evaluation demonstrates that this architecture reduces the FNR by over 30 percentage points compared to state-of-the-art baselines and maintains robust security coverage even under operational distribution shifts.

\printbibliography

\appendices

\section*{Appendix A: Additional Details About NetInspector-Bench}
In this section, we provide more details of our benchmark. NetInspector-Bench extends and reframes prior conflict detection datasets~\cite{jacobs2021hey} in two ways. First, the original Jacobs et al.\ dataset covers only campus-scale networks with a limited set of syntactic conflict patterns. We expand coverage to three network domains (Campus, Enterprise, WAN) and introduce new violation scenarios that require multi-hop topological reasoning and hierarchical group resolution — scenarios that the original benchmark does not test. Second, and more importantly, we reframe the task: rather than treating the six violation types as abstract ``conflict classes,'' we map them to three concrete security threat classes — Privilege Escalation (PE: negation, hierarchical, synonym), Isolation Bypass (IB: path), and Denial of Service (DoS: time, QoS) — enabling security-oriented evaluation of enforcement systems.

Table~\ref{tab:mini-bench} provides a full breakdown. The dataset spans 2,224 samples across the three topologies. To reflect realistic deployment conditions, only $\sim$20\% of samples (437) are security-violating intents and 80\% (1,787) are benign — a class imbalance that challenges systems to maintain high sensitivity without generating excessive false alarms. The threat class distribution across all topologies is: 173 PE violations, 87 IB violations, and 140 DoS violations.

\begin{table}[h]
\centering
\caption{Fine-tuning parameters and metrics\label{tab:fine_tuning}}
\normalsize  
\setlength{\tabcolsep}{6pt}  
\begin{tabular}{ll}
\hline
\textbf{Parameter} & \textbf{Value} \\
\hline
Trained tokens & 1,709,187 \\
Epochs & 3 \\
Batch size & 2 \\
LR multiplier & 2 \\
Training time & 30 minutes \\
Training cost & \$15 \\
\hline
\end{tabular}
\end{table}

\begin{table*}[t]
\centering
\caption{NetInspector-Bench breakdown. Six violation types map to three security threat classes: Privilege Escalation (PE) $=$ negation $+$ hierarchical $+$ synonym; Isolation Bypass (IB) $=$ path; Denial of Service (DoS) $=$ time $+$ qos.}
\label{tab:mini-bench}
\begin{tabular}{lcccccccccccc}
\hline
\textbf{Topology} & \textbf{\#Samples} & \textbf{\#Violation} & \textbf{\#Benign}
& & \multicolumn{6}{c}{\textbf{Violation Type}}
& & \multicolumn{1}{c}{\textbf{Threat Class}} \\
\cline{6-11} \cline{13-13}
 &  &  &  &  & \textbf{negation} & \textbf{hierarchical} & \textbf{synonym} & \textbf{path} & \textbf{time} & \textbf{qos}
 & & \textbf{PE / IB / DoS} \\
\hline
Campus     & 1224 & 237  & 987  & & 37 & 33 & 28 & 42 & 30 & 30 & & 98 / 42 / 60 \\
Enterprise & 500  & 100  & 400  & & 5  & 15 & 10 & 25 & 10 & 35 & & 30 / 25 / 45 \\
WAN        & 500  & 100  & 400  & & 15 & 15 & 15 & 20 & 10 & 25 & & 45 / 20 / 35 \\
\hline
Total      & 2224 & 437  & 1787 & & 57 & 63 & 53 & 87 & 50 & 90 & & 173 / 87 / 140 \\
\hline
\end{tabular}
\end{table*}

\section*{Appendix B: Prompt Engineering and Fine-Tuning Configuration}

To comprehensively evaluate the capabilities of Large Language Models in network intent translation and security invariant enforcement, we designed a set of prompt templates ranging from simple instructions to complex, context-aware structures. As detailed in Table~\ref{tab:prompt_templates}, our experimental design covers three primary strategies:

\begin{itemize}
    \item \textbf{Zero-Shot:} This baseline approach provides the model with a concise task description and the input query directly, testing the model's innate ability to generalize without prior examples.
    \item \textbf{Few-Shot:} We augment the zero-shot template by prepending multiple randomly selected demonstration examples. This in-context learning strategy aims to guide the model toward the expected output format and logic by providing immediate references.
    \item \textbf{Few-Shot w/ RAG:} To address the complexity of domain-specific security invariant reasoning, we implemented a hybrid Retrieval-Augmented Generation (RAG) template. This approach establishes a system role of ``network intent translator'' and dynamically retrieves semantically relevant examples using the \textit{text-embedding-3-large} model with a Maximal Marginal Relevance (MMRE) retrieval strategy. The final user prompt incorporates the top-$k$ retrieved examples—ranked by cosine similarity and filtered for contextual relevance—alongside domain-specific guidelines and entity-annotated queries.
\end{itemize}

Complementing our prompt engineering efforts, we also evaluated the efficacy of fine-tuning. We adopted a lightweight training configuration to balance performance gains with computational efficiency. As summarized in Table~\ref{tab:fine_tuning}, the model was trained on a dataset of 1,709,187 tokens over 3 epochs. We utilized a batch size of 2 and a learning rate multiplier of 2. The entire fine-tuning process was notably efficient, completing in just 30 minutes with a total compute cost of \$15. This low resource footprint demonstrates the feasibility of adapting frontier models for specialized network tasks without prohibitive infrastructure requirements.

\newcolumntype{L}[1]{>{\raggedright\arraybackslash}p{#1}}
\begin{table*}[t]
\centering
\normalsize
\begin{tabular}{@{}L{2.5cm} L{6.5cm} L{6.5cm}@{}}
\toprule
\textbf{Template Name} & \textbf{Description} & \textbf{Template} \\
\midrule
0-shot & Describes the task and directly gives the input query. &
\textbf{USER} $\langle$task description$\rangle$ $\langle$input$\rangle$ \\
\midrule
few-shot & Describes the task and provides multiple random-selected demonstration examples before the query. &
\textbf{USER} $\langle$task description$\rangle$ $\langle$demonstration example 1$\rangle$ $\langle$demonstration example 2$\rangle$ $\langle$demonstration example 3$\rangle$ $\langle$demonstration example 4$\rangle$ $\langle$input$\rangle$ \\
\midrule
Few-shot w/ RAG & employs a hybrid retrieval-augmented generation approach, where semantically relevant examples are dynamically retrieved from a vector database based on input query embedding similarity. The retrieved examples are ranked by cosine similarity and filtered to maximize contextual relevance while maintaining prompt diversity. &
\begin{minipage}[t]{6.5cm}
\textbf{SYSTEM} $\langle$role: network intent translator$\rangle$ \\ $\langle$context: embedding model=text-embedding-3-large, retrieval strategy=MMRE$\rangle$

\textbf{USER} $\langle$top-$k$ semantically retrieved examples with similarity scores$\rangle$ $\langle$domain-specific guidelines$\rangle$ $\langle$task description$\rangle$ $\langle$query with entity annotations$\rangle$
\end{minipage} \\
\bottomrule
\end{tabular}
\caption{Templates for system prompts used in the measurement study.}
\label{tab:prompt_templates}
\end{table*}

\begin{figure}[ht]
	\centering
	\includegraphics[width=0.5\textwidth]{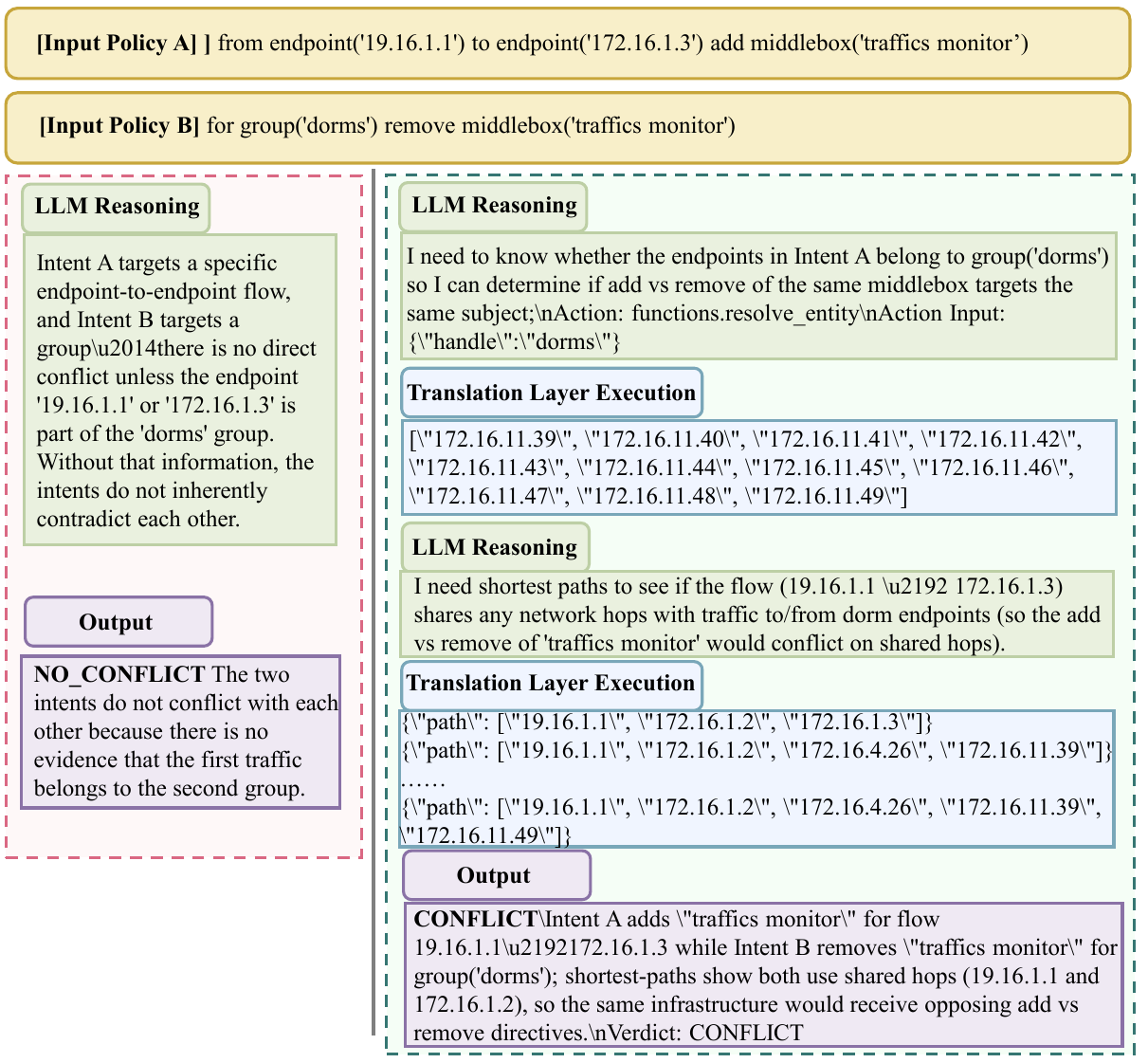}
	\caption{Summarized example of a multi-step reasoning process to mitigate Privilege Escalation. The left shows the execution of LLM reasoning without NetInspector, while the right showcases the multi-step execution of LLM with NetInspector, including more than one round of interaction between the LLM and NetInspector.}\label{fig:case-study-muliti-step}
\end{figure}

\section*{Appendix C: Examples of LLM Failures}

\begin{figure*}[ht]
	\centering
	\includegraphics[width=0.95\textwidth]{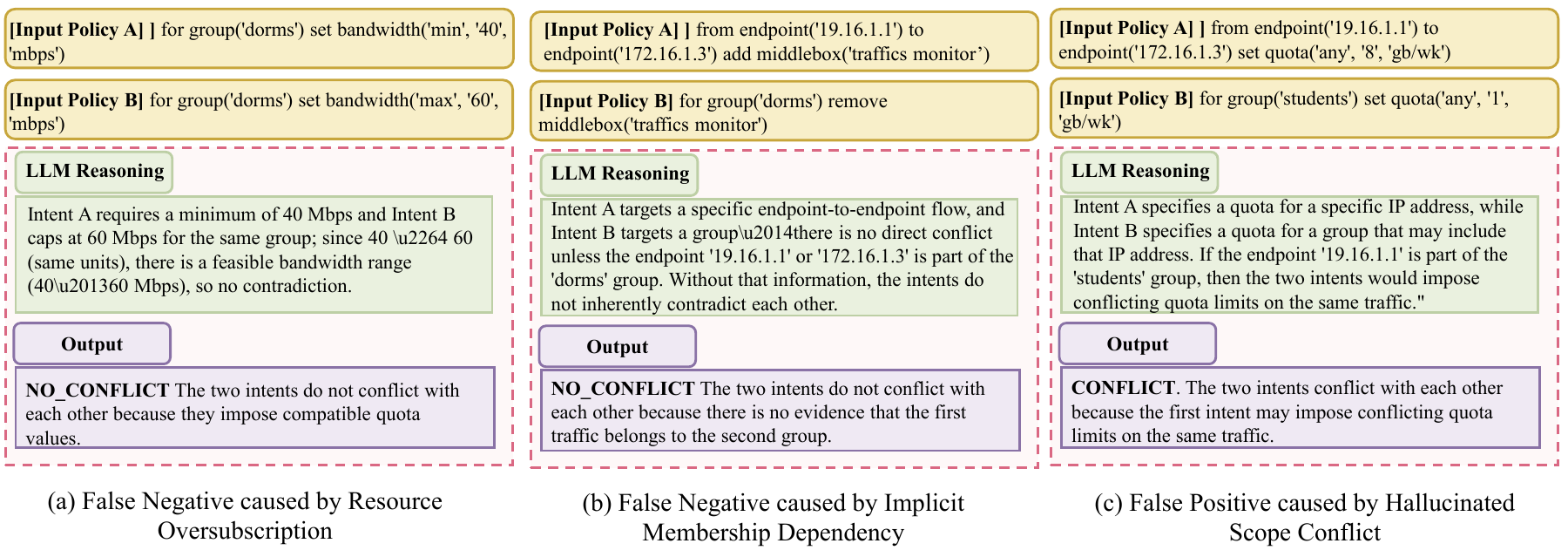}
	\caption{Summarized example of cases where naive LLM-based solutions fail.}\label{fig:example-failed-cases}
\end{figure*}

Figure~\ref{fig:example-failed-cases} illustrates three critical failure modes where naive LLM-based solutions struggle due to a lack of domain grounding. First, Figure~\ref{fig:example-failed-cases}(a) demonstrates a \textit{False Negative caused by Resource Oversubscription}. In this scenario, the model analyzes two bandwidth intents---one setting a minimum guarantee of 40 Mbps and another setting a maximum cap of 60 Mbps. The LLM incorrectly outputs \texttt{NO\_CONFLICT} by relying solely on abstract numerical logic (reasoning that $40 \leq 60$ creates a valid range), thereby failing to account for physical network constraints where satisfying the minimum guarantee might be impossible due to capacity oversubscription. Second, Figure~\ref{fig:example-failed-cases}(b) depicts a \textit{False Negative caused by Implicit Membership Dependency}. Here, one intent applies to a specific endpoint (19.16.1.1) while the conflicting intent applies to a logical group ('dorms'). Lacking access to the underlying network state or directory services, the LLM fails to resolve whether the endpoint belongs to the group. It treats the scopes as disjoint and erroneously predicts \texttt{NO\_CONFLICT}, missing the contradiction inherent in the endpoint's actual membership. Conversely, Figure~\ref{fig:example-failed-cases}(c) highlights a \textit{False Positive caused by Hallucinated Scope Conflict}. The reasoning that an endpoint \textit{might} be part of a group leads the model to hallucinate an overlap, resulting in an unjustified \texttt{CONFLICT} prediction despite the absence of evidence linking the two scopes.

\section*{Appendix D: Additional Case Study}
\parhead{Case Study 4: Multi-Step Reasoning to Prevent Security Regression}
In this seciton we present another example of how NetInspector detects potential Privilege Escalation. In this example, the input intents actually requires NetInspector to perform multi-step reasoning to identify the security violation. Figure~\ref{fig:case-study-muliti-step} illustrates a sophisticated defense against unintentional Privilege Escalation, where broad administrative actions inadvertently dismantle granular security controls. In this scenario, the system must reconcile opposing operational semantics across different abstraction levels: one intent explicitly \textit{adds} a traffic monitor middlebox to a specific flow, while a second intent \textit{removes} the same middlebox for an entire user group. A naive LLM, struggling with both hierarchical set containment and topological physicalities, perceives these as non-conflicting, distinct tasks. This oversight creates a dangerous vulnerability where a high-level maintenance command could silently disable essential monitoring for sensitive flows.
NetInspector resolves this ambiguity through a multi-step grounding process that chains entity resolution with topological verification. First, the agent invokes the Translation Layer to verify group membership, confirming that the specific flow is indeed a subset of the target group. Second, it retrieves the shortest path data to determine if the directives physically converge on the same infrastructure. By confirming that the \textit{add} and \textit{remove} operations target the same shared hops, the LLM correctly identifies the security invariant violation. This capability to chain reasoning steps is crucial for maintaining security coverage in complex environments, ensuring that broad policy changes do not accidentally regress the security posture of specific critical assets.

\end{document}